\documentclass{emulateapj}
\usepackage{apjfonts}
\journalinfo{The Astrophysical Journal, 2010, in press}

\shorttitle{SOLAR WIND VIA MAGNETIC CARPET RECONNECTION?}
\shortauthors{CRANMER AND VAN BALLEGOOIJEN}

\begin{document}

\title{Can the Solar Wind be Driven by Magnetic Reconnection in
the Sun's Magnetic Carpet?}

\author{Steven R. Cranmer and Adriaan A. van Ballegooijen}
\affil{Harvard-Smithsonian Center for Astrophysics,
60 Garden Street, Cambridge, MA 02138}
%\email{scranmer@cfa.harvard.edu,avanballegooijen@cfa.harvard.edu}
\email{scranmer@cfa.harvard.edu}

\begin{abstract}
The physical processes that heat the solar corona and accelerate
the solar wind remain unknown after many years of study.
Some have suggested that the wind is driven by waves and turbulence
in open magnetic flux tubes, and others have suggested that plasma
is injected into the open tubes by magnetic reconnection with closed
loops.
In order to test the latter idea, we developed Monte Carlo
simulations of the photospheric ``magnetic carpet'' and extrapolated
the time-varying coronal field.
These models were constructed for a range of different magnetic
flux imbalance ratios.
Completely balanced models represent quiet regions on the Sun and
source regions of slow solar wind streams.
Highly imbalanced models represent coronal holes and source regions
of fast wind streams.
The models agree with observed emergence rates, surface flux
densities, and number distributions of magnetic elements.
Despite having no imposed supergranular motions in the models,
a realistic network of magnetic ``funnels'' appeared spontaneously.
We computed the rate at which closed field lines open up (i.e.,
recycling times for open flux), and we estimated the energy
flux released in reconnection events involving the opening up of
closed flux tubes.
For quiet regions and mixed-polarity coronal holes, these energy
fluxes were found to be much lower than required to accelerate
the solar wind.
For the most imbalanced coronal holes, the energy fluxes may be
large enough to power the solar wind, but the recycling times are
far longer than the time it takes the solar wind to accelerate
into the low corona.
Thus, it is unlikely that either the slow or fast solar wind is
driven by reconnection and loop-opening processes in the magnetic
carpet.
\end{abstract}


\section{Introduction}
\label{sec:1}

The magnetic field in the solar photosphere exists in a complex and
continually evolving state that is driven by convective motions
under the surface.
The dynamic interplay between the magnetic field and the plasma
has been called the Sun's ``magnetic carpet'' \citep{TS98}.
There is a clear correlation between the topology and strength
of the magnetic field and the energy deposition that is responsible
for the hot ($T \gtrsim 10^{6}$ K) solar corona.
We also know that the gas pressure associated with coronal heating
is an important contributor to accelerating the supersonic solar
wind \citep{Pk58}.
Thus, it is natural to wonder to what extent the magnetohydrodynamic
(MHD) motions in the magnetic carpet are ultimately responsible
for producing at least some of the solar wind's mass loss.

Recently, two distinct classes of theoretical explanation have been
proposed for the combined problem of coronal heating and solar wind
acceleration.
In the {\em wave/turbulence-driven} (WTD) models, convection
jostles the open magnetic flux tubes that are rooted in the
photosphere and produces waves that propagate into the corona.
These waves (usually assumed to be Alfv\'{e}n waves) are proposed
to partially reflect back down toward the Sun, develop into MHD
turbulence, and heat the plasma by their gradual dissipation
\citep{Ho86,Ve91,WS91,Mt99,SI06,CvB07,Wa09,VV10,MS10}.
In the {\em reconnection/loop-opening} (RLO) class of models,
it is assumed that closed, loop-like magnetic flux systems are
the dominant source of mass and energy into the open-field regions.
Some have suggested that RLO-type energy exchange primarily occurs
on small, supergranular scales \citep{Ax92,Fi99,Fi03,SM03}.
However, other models have been proposed in which the
``interchange reconnection'' occurs in and between large-scale
coronal streamers further from the solar surface
\citep{Ei99,SN04,An10}.

The WTD idea of a flux tube that is open---and which stays
open as the wind accelerates---is conceptually simpler than the
idea of frequent changes in the flux tube topology.
Because of this simplicity, the WTD models have been subject to a
greater degree of development and testing than the RLO models.
In addition, we have a great deal of observational evidence that
waves and turbulent motions are present everywhere from the
photosphere to the heliosphere
\citep[see, e.g.,][]{TM95,BC05,Hn07,As08}.
Thus, it is of interest to pursue the WTD idea to see how these
waves affect the mean state of the plasma in the absence of any
other sources of energy.
For example, \citet{CvB07} and \citet{Cr09} showed that a set of
WTD models that varied only the magnetic flux-tube expansion rate
(and kept all other parameters fixed, including the wave fluxes at
the lower boundary) can successfully predict a wide range of
measured properties of both fast and slow solar wind streams.

RLO models need to be subjected to the same degree of development,
testing, and refinement as the WTD models.
This idea has a natural appeal since the open flux tubes must be
rooted in the vicinity of closed loops \citep{Do86}.
In fact, multiple RLO-like reconnection events have been observed
in coronal holes as ``polar jets'' by instruments aboard
{\em SOHO,} {\em Hinode,} and {\em STEREO}
\citep[e.g.,][]{We98,So07,Ns09}.
Reconnection at the edges of coronal holes may be necessary to
produce their observed rigid rotation \citep{Li06}.
There are also observed correlations between the lengths of coronal
loops, the electron temperature in the low corona, and the wind
speed in interplanetary space \citep{Gl03} that are highly suggestive
of a net transfer of magnetic energy from the loops to the
open-field regions \citep[see also][]{Fi99,Fi03}.

Testing the RLO idea using theoretical models is more difficult than
testing the WTD idea because of the complex multi-scale nature
of the relevant magnetic fields.
Many aspects of RLO-type processes cannot be simulated without
resorting to fully three-dimensional and time-dependent models of
the connection between the magnetic carpet and the solar wind.
The goal of this paper is to begin constructing such models in order
to address several of the following unanswered questions about
the RLO model.
For example, how much of the magnetic energy that is liberated by
reconnection goes into simply reconfiguring the closed fields, and
how much goes into changing closed fields into open fields?
Specifically, what is the actual rate at which magnetic flux
opens up from the magnetic carpet?
Can the observed polar jets provide enough energy to drive a
significant fraction of the solar wind?
Lastly, how is the reconnection energy distributed into various
forms (e.g., bulk kinetic energy, thermal energy, waves, or
energetic particles) that can each affect the accelerating wind in
different ways?

In this paper we present Monte Carlo models of the solar magnetic
carpet that are used to determine the topology, temporal variability,
and energy flux along field lines connected with the accelerating
solar wind.
Section \ref{sec:2} gives an overview of the motivations behind our
choices of modeling technique.
In Section \ref{sec:3} we describe the physical ingredients that
went into the Monte Carlo models of the photospheric magnetic field.
Section \ref{sec:4} then presents the results of these models and
compares them with a range of observational diagnostics.
In Section \ref{sec:5} we then describe how field lines were
extrapolated from the photospheric lower boundary up into the corona,
and we discuss the resulting time scales and energy fluxes that were
derived for flux tubes relevant to RLO wind acceleration models.
Finally, Section \ref{sec:6} concludes this paper with a brief
summary of the major results, a discussion of some of the wider
implications of this work, and suggestions for future improvements.

\section{Motivations and Methods}
\label{sec:2}

In this section we summarize the techniques that we chose to
simulate the connections between the photospheric magnetic field
and the open flux tubes feeding the solar wind.
It is also important to clarify how and why our assumptions are
consistent with the goal to quantify the impact of RLO physical
processes.
Our modeling was done in two steps.
First, we simulated the photospheric magnetic carpet by means of a
Monte Carlo ensemble of positive and negative monopole sources of
magnetic flux.
These sources are assumed to emerge from below (as bipolar ephemeral
regions), move around on the surface, merge or cancel with their
neighbors, and spontaneously fragment.
We specified the rates and other details about these processes by
comparing with many different observational constraints.
Second, we used the photospheric flux sources to extrapolate field
lines up into the corona by assuming a potential field.

Despite the model's reliance on flux emergence from below the solar
surface, we did not model the subphotospheric motions explicitly.
A complete treatment of this problem should describe how the
photospheric fields are ultimately controlled by the overturning
dynamics of convection cells and their interactions with one
another \citep[e.g.,][]{Fg10,St10}.
In many ways, however, the photosphere is believed to act as a
relatively ``clean'' transition layer between the highly fragmented
fibril fields of the convection zone and the space-filling fields
of the corona \citep{Am05,vBM07}.
We take advantage of the rapid change in plasma conditions between
these regions to utilize the thin photospheric layer as a natural
lower boundary.
Thus, we used observations of individual features and their motions
to set up statistical rules for how these features evolve in our
Monte Carlo models of the photosphere.
The ultimate test of the validity of these rules is that the
resulting complex and multi-scale photospheric field matches a
wide range of observations.
(Of course, the observations used to test the models must be
independent of the observations that were used to determine the
rules; see Section \ref{sec:4} below for more details.)

Many earlier studies of magnetic flux transport in the photosphere
were focused on the net horizontal diffusion of fields
\citep[e.g.,][]{Wa89,Si95,vB98}.
A new era was ushered in by \citet{Sj97}, who constructed a
statistical model that also included flux emergence, cancellation,
merging, and fragmenting.
Numerical simulations of these effects were also produced by
\citet{Pn01}, \citet{Si01}, and \citet{Ch07}.
Our Monte Carlo models of the photospheric magnetic carpet are based
on these earlier models, but with three main differences:
(1) we use more up-to-date flux emergence rates \citep{Hg08,Hg10},
which give at least an order of magnitude faster ``recycling time''
for photospheric flux;
(2) we model both balanced and imbalanced regions on the solar
surface that are designed to simulate both quiet Sun and coronal
hole areas; and
(3) we do not presume the existence of supergranular motions on the
surface---but the model does produce a network-like organization
of the field as a natural output \citep[e.g.,][]{Ra03}.

At each time step in the Monte Carlo simulations, we extrapolate
magnetic field lines up into the corona by assuming the field is
derivable from a scalar potential.
Although the actual solar field is likely to have significant
non-potential components \citep[e.g.,][]{Sd09,Ed09}, the approximation
of a potential field has been found to be useful in identifying the
regions where magnetic reconnection must be taking place
\citep{Lg96,Cl05}.
The potential-field method is also many orders of magnitude more
computationally efficient than solving the full three-dimensional
MHD conservation equations.
(Doing the latter for a system with a complex, evolving,
magnetic-carpet-like lower boundary is still prohibitively
expensive in terms of computation time.)
Our method involves ignoring the ``internal'' details about how
magnetic reconnection actually affects the coronal plasma and only
investigating the magnetic energy that is lost via reconnection.
We use Longcope's (\citeyear{Lg96}) {\em minimum current corona}
model to take account of the reconnection energetics.
We emphasize that---despite the title of this paper---magnetic
reconnection is not a primary ``driver'' unto itself and is merely
the end product of the flux emergence, cancellation, merging,
fragmentation, and diffusion that occurs on the photospheric lower
boundary.

By modeling only the net changes in the magnetic field from
one time step to the next, we end up ignoring some potentially
important plasma effects.
For example, \citet{PG04} showed that reconnection may progress
much more slowly in full MHD than one would expect from modeling
the system as an idealized succession of potential-field states.
Also, \citet{Ly08}, \citet{Pt09}, \citet{Ed10a}, and others have
shown that long-lived, field-aligned currents can exist in the
corona due to the injection of magnetic flux from below, and these
energetically important structures are not accounted for in
potential-field models.
However, we do not model the most topologically complicated
regions of the corona, such as the footpoints of field lines that
connect to the cusps of helmet streamers, or to the heliospheric current
sheet, or to other large-scale separatrix and quasi-separatrix layers
\citep[see, e.g.,][]{Ed09,An10}.
Our models generally presume the existence of a simple unipolar
field at a large height, in conjunction with the complex
and time-varying magnetic carpet field at the bottom.
These ``open'' unipolar fields may in fact close back down onto the
solar surface on spatial scales larger than our modeled patches of
the Sun.
Whether this occurs or not depends on the global distribution of
magnetic flux across the entire solar surface, which is beyond the
scope of this paper to model.

There have been many three-dimensional MHD simulations of the
coronal response to underlying photospheric motions 
\citep[see also][]{GN05,Ph06,Ga06,Ib08},
and this paper does not attempt to reproduce those results.
The spatial and temporal complexity of the footpoint motions in
most MHD models, however, has usually been assumed to be simpler
than in the full magnetic carpet as modeled here.
We also ignore the possibility that there could be a significant
back-reaction from the corona on the dynamics of the photospheric
footpoints \citep[see][]{Gr08}.
Others have studied how the evolving photospheric field can affect
the properties of coronal Alfv\'{e}n waves \citep{Ma07},
coronal mass ejections \citep{Ly09,Ye10}, and the large-scale
heliospheric magnetic field \citep{Ji10}.
The goal of this paper is much more limited.
We aim to take an initial census of the rate at which closed
flux opens up from the Sun's magnetic carpet, and to estimate how
much magnetic energy may be released by the attendant reconnection.
Thus, this paper is envisioned as a kind of ``pathfinder'' study
that carves out the order-of-magnitude expectations for what more
sophisticated MHD simulations are likely to reveal in detail.

\section{Photospheric Field Evolution: Model}
\label{sec:3}

In our model, the topology and energy balance of the coronal
magnetic field are assumed to be fully determined by the lower
boundary conditions at the solar photosphere.
Here we describe how the photospheric field can be simulated by
assuming it consists of a collection of evolving flux sources.
We developed a FORTRAN code called BONES to produce Monte Carlo
simulations of these flux sources and to trace magnetic flux tubes
up into the corona.
The title BONES was inspired by the popular conception of the solar
magnetic field as a topological {\em skeleton} for locating
important sites of energy release \citep{Pn08}, and also by the
dependence on randomness in the Monte Carlo technique (i.e.,
``rolling the bones'').

For a Monte Carlo simulation like this, it is not possible to write
down a single set of equations that governs the behavior of the
magnetic field.
Each simulation is a particular realization of an ensemble of
possible states \citep[see also][]{Sj97}.
Therefore, we must describe the individual processes that govern
the motion and evolution of the flux elements.
Section \ref{sec:3.1} introduces some of the general attributes
of the BONES simulations.
The code models the time dependence of the photospheric field as
the net result of four processes: emergence of new bipoles
(Section \ref{sec:3.2}), random horizontal motions
(Section \ref{sec:3.3}), merging and cancellation between pairs
of nearby elements (Section \ref{sec:3.4}), and spontaneous
fragmentation (Section \ref{sec:3.5}).

\subsection{Basic Properties and Initial Conditions}
\label{sec:3.1}

We modeled a patch of the photospheric solar surface as a horizontal
square box that extends 200 Mm on each side.
This length scale was chosen to be large enough to encompass several
supergranular network cells, but small enough to be applicable to
solar wind source regions of roughly uniform character (i.e.,
coronal holes or quiet Sun) and to be able to ignore the radial
curvature of the solar surface.
Thus, the surface area of the model domain is defined as
$A = 4 \times 10^{20}$ cm$^2$, or about 0.7\% of the Sun's surface
area.

In the part of the BONES code that evolves the photospheric magnetic
field, each flux element is considered to be a point-like monopole
having only three attributes: an $x$ position, a $y$ position,
and a signed magnetic flux $\Phi$.
Even though many elements are injected into the simulation in
equal-and-opposite pairs (i.e., as the footpoints of bipole loops), the
code retains no memory of that association in subsequent time steps.
We quantized the magnetic flux in units of $10^{17}$ Mx so that
incomplete cancellations do not produce a huge number of
infinitesimally small elements \citep[see, e.g.,][]{Pn01}.

We computed the continuous magnetic field that results from the
flux elements in several ways.
In Section \ref{sec:5.1} we describe the computation of the vector
field ${\bf B}$ above the photospheric surface.
Here we show how an upper limit on the magnetic field
strength in the flux elements (in the photosphere) can be used to
obtain a lower limit on their spatial extent.
Let us assume that the horizontal cross section of a flux element is
circular, and that it is filled with a constant vertical magnetic field.
It is generally assumed that the field in small photospheric
concentrations cannot be significantly stronger than the so-called
equipartition field, in which the plasma is in total pressure
equilibrium with its (approximately field-free) surroundings.
In this case, the upper limit on the field strength is
$B_{\rm max} \approx 1400$ G \citep[see, e.g.,][]{Pk76,Li02,CvB05}.
Thus, we can estimate a lower limit to the radius of the circular
flux element as
\begin{equation}
  r_{c} \, = \, \sqrt{\frac{|\Phi|}{\pi B_{\rm max}}}  \,\, .
  \label{eq:rc}
\end{equation}
The typical size of observed intergranular G-band bright points is
$r_{c} \approx 50$--150 km \citep{MK83}.
Recently, \citet{SA10} measured the filling factor ($f = 0.89$\%) and
number density ($\rho = 0.97$ Mm$^{-2}$) of bright points in quiet
Sun regions, and these values are consistent with a radius of
$r_{c} = (f/ \pi\rho)^{1/2} \approx 55$ km.
The above range of sizes corresponds appropriately to fluxes at
the low end of the range simulated here; i.e., between
$10^{17}$ and $10^{18}$ Mx.
Elements with larger fluxes may not be completely filled by
equipartition fields, and thus they would have larger spatial extents
than expected from Equation (\ref{eq:rc}).

At any one time in the simulation, the sum of all positive fluxes
is denoted $\Phi_{+}$ and the sum of all negative fluxes is denoted
$\Phi_{-}$.
These are signed quantities, with $\Phi_{+} > 0$ and $\Phi_{-} < 0$.
For all models discussed below that have an imbalance between the two
polarities, the sense of the imbalance is always to have
$|\Phi_{+}| > |\Phi_{-}|$.
All results should be equivalent for imbalances in the opposite sense.
The mean magnetic flux densities in the positive and negative
flux elements, taken over the entire simulation domain, are denoted
$B_{\pm} = \Phi_{\pm} / A$.
Thus, the total ``unsigned'' or absolute flux density is given by
$B_{\rm abs} = B_{+} + |B_{-}|$ and the net flux density is given by
$B_{\rm net} = |B_{+} + B_{-}| = B_{+} - |B_{-}|$.
The simulation's flux imbalance fraction $\xi$ is defined as
$\xi = B_{\rm net}/B_{\rm abs}$.
Small values for this ratio (i.e., $\xi \lesssim 0.3$) are typical
for quiet Sun regions, and larger values ($\xi \gtrsim 0.7$) are
typical for coronal holes \citep{WS04,Zh06,Hg08,Ab09}.

Each run of the BONES code begins with specified initial conditions
at time $t = 0$.
For models having $\xi = 0$, there are no flux elements in the
domain at the beginning of the simulation.
Perfect flux balance is maintained by having all new flux elements
emerge into the domain at later times as balanced bipoles.
For models having $\xi > 0$, the simulation begins with a number
of identical flux elements, all having positive polarity, that are
distributed randomly over the surface $A$.
These initial elements are assumed to each have an equal flux
given by 0.1 times the mean flux in an emerging bipole (see
Section \ref{sec:3.2}).
The number of these initial elements is determined by the input
value of the net flux density $B_{\rm net}$.
As in the $\xi = 0$ case, all new flux elements that enter the
domain at $t > 0$ are balanced pairs, and thus $B_{\rm net}$ remains
exactly constant as a function of time.

For a given simulation that is intended to model a patch of the Sun
having an imposed flux imbalance ratio $\xi$, the choice of the
proper input value of $B_{\rm net}$ is not known at the outset.
The overall level of magnetic flux that ends up existing in the
simulation depends on the collection of dynamical parameters that
describe the flux emergence, fragmentation, horizontal diffusion,
and merging (see below).
Specifically, the emergence rate $E$ depends explicitly on $\xi$
\citep[e.g.,][]{Hg08}.
Thus, for a given set of dynamical parameters and a desired value of
$\xi$, we had to produce an iterative set of trial runs with a
range of guesses for $B_{\rm net}$.
Only one unique value of $B_{\rm net}$ gave rise to a model having
the proper self-consistent value of $\xi$.
After doing this for a range of models, the relationship between
these two parameters was fit with the following approximate
relation,
\begin{equation}
  \xi \, \approx \, \frac{0.268 \, B_{\rm net}}
  {[1 + (B_{\rm net} / 3.58)^{2.71}]^{0.365}}  \,\, ,
  \label{eq:xifit}
\end{equation}
where $B_{\rm net} > 0$ is measured in Gauss and $\xi$ is
dimensionless.

The discrete time step chosen for the simulations was
$\Delta t = 300$ s, the same as that used by \citet{Pn01}.
Five minutes is a representative time scale for photospheric
granulation \citep[e.g.,][]{DG84},
so using a smaller time scale would only be appropriate
if the coherent granular motions were being modeled explicitly.
\citet{AR09} found that on spatial scales longer than 300--500 km
the solar granulation acts as a stochastic, Markovian process.
For representative granulation velocities of order 1 km s$^{-1}$
\citep{Hz02},
this confirms that the minimum resolvable time scale (when ignoring
coherent convective overturning) should be about 300--500 s.
For all processes in the BONES code that are simulated as occurring
stochastically, we used the RAN2 random number generator of
\citet{Pr92}.
This routine does not repeat its pseudo-random sequence until
called at least $2 \times 10^{18}$ times.
This limit was never approached, since in even the longest runs of
the code the RAN2 routine was never called more than $10^{10}$ times.

Over the course of each time step $\Delta t$, the code updates the
properties of each of the flux elements from the effects of the
four general sets of processes described below.

\subsection{Flux Emergence}
\label{sec:3.2}

Bipolar magnetic features are observed to emerge from beneath the
photosphere with fluxes spanning several orders of magnitude from
$\sim$10$^{16}$ Mx (internetwork concentrations) to 
$\sim$10$^{22}$ Mx (sunspots) \citep{Sj01,Pn02,Hg08}.
Away from active regions, much of the emergence tends to occur in
the form of bipolar {\em ephemeral regions} (ERs) with
$|\Phi| \approx 10^{18}$--$10^{19}$ Mx
\citep[see, e.g.,][]{HM73}.
The individual poles of ERs often are advected to the edges of
supergranular cells and coalesce to form {\em network
concentrations} that end up with similar absolute fluxes as the
ERs themselves \citep{Ma88}.

The rate of emergence of ER flux, which we denote $E$, has been
estimated in various ways from both measurements and models.
As the sensitivity and cadence of observations has improved, the
derived emergence rates have generally increased.
\citet{Sj01} reviewed earlier measurements and models that pointed
to a range of $E$ values between about $2 \times 10^{-6}$ and
$4 \times 10^{-5}$ Mx cm$^{-2}$ s$^{-1}$.
Earlier Monte Carlo models also found that values in this range
seemed to behave in similar ways as the real Sun.
For example, \citet{Pn01} used $E \approx 8 \times 10^{-6}$
Mx cm$^{-2}$ s$^{-1}$, and \citet{Si01} used
$E \approx 1.3 \times 10^{-5}$ Mx cm$^{-2}$ s$^{-1}$.
\citet{KR03} found that a slightly higher value of
$9 \times 10^{-5}$ Mx cm$^{-2}$ s$^{-1}$ was needed to reproduce
{\em TRACE} measurements of the chromospheric network.
Assuming a mean flux density in the quiet Sun of about 3 to 4
Mx cm$^{-2}$, it is possible to use the above emergence rates to
estimate ``flux recycling times'' between about 0.5 and 20 days.

However, many of these earlier measurements were made with sequences
of relatively low-cadence magnetograms.
\citet{Hg08} found that when the cadences are reduced from about
90 min to 5 min, many more emergence events are observed and the
emergence rate increases.
In fact, \citet{Ma88} claimed that it is virtually impossible to
even {\em identify} the same ER from one image to the next unless
the time cadence between them is shorter than about 10 min.
The revised analysis of \citet{Hg08} showed that values as large
as $E \approx 10^{-3}$ Mx cm$^{-2}$ s$^{-1}$ are often seen in
regions of balanced magnetic polarities,\footnote{%
Figure 5 of \citet{Hg08} showed values that were erroneously reduced
in magnitude.
The values given in Table 2 of \citet{Hg08} represented the correct
magnitudes for the emergence rates, and a corrected revision of
their Figure 5 was presented by \citet{Hg10}.
Our fits to these observations utilized a multiplicative
correction factor of 5 to the numbers shown in their original
Figure 5(b), which is consistent with the updated version shown by
\citet{Hg10}.}
along with a noticeable decrease in $E$ as $\xi$ increases from 0 to 1.
For most values of the imbalance ratio ($\xi \lesssim 0.8$),
these rates of emergence are consistent with flux recycling times
of only 1--2 hr.

We fit the modified rates shown in Table 2 and Figure 5 of
\citet{Hg08,Hg10} with a quadratic function of the imbalance
ratio $\xi$, and found
\begin{equation}
  E \, = \, 7.928 \times 10^{-4} \left( 1.356 - \xi^{2} \right)
  \,\,\,\, \mbox{Mx} \,\, \mbox{cm}^{-2} \,\, \mbox{s}^{-1}
  \,\, .
  \label{eq:Efit}
\end{equation}
For a region with balanced magnetic flux ($\xi=0$), the maximum
value of the emergence rate is $E = 1.075 \times 10^{-3}$
Mx cm$^{-2}$ s$^{-1}$.
As $\xi \rightarrow 1$, the parameterized rate declines to a
minimum value of $E = 2.824 \times 10^{-4}$ Mx cm$^{-2}$ s$^{-1}$.
Note, however, that the largest imbalance fraction in the
measurements of \citet{Hg08} was $\xi \approx 0.94$.
Our use of values larger than this represents extrapolation.
It is possible that $E$ may decrease more rapidly---possibly
to zero---as $\xi$ increases from 0.94 to 1.
In any case, we never model the completely unipolar case of
$\xi=1$.
The largest value of $\xi$ used in the models presented
below is 0.99.

In order to determine the number of bipoles ($N_{\rm em}$) that
emerge in each time step in the simulation domain, we adopted a
fiducial value for the average flux per bipole,
$\langle \Phi \rangle = 9 \times 10^{18}$ Mx (see below).
Thus, $N_{\rm em} = E A \Delta t / \langle \Phi \rangle$.
In general, this does not yield an integer number of bipoles.
For a given non-integer value of $N_{\rm em}$ that falls between
the two integers $n$ and $n+1$, we used the fractional remainder
of $N_{\rm em}$ (in excess of $n$) to determine the statistical
chance that the resulting number of bipoles is either $n$ or
$n+1$.
For example, if $N_{\rm em} = 10.22$, there is a 22\% chance
that there will be 11 bipoles, and a 78\% chance there will be 10
bipoles.
A new random number is generated in each time step to determine
whether there will be $n$ or $n+1$ new bipoles.

For each of the emerging bipoles, the BONES code determines its
total absolute flux by drawing from an empirically constrained
probability distribution of the form
\begin{equation}
  P_{\rm E} (\Phi)  =  \left\{
  \begin{array}{ll}
    (\Phi - \Phi_{\rm min}) \exp \left[ - (\Phi - \Phi_{\rm min})
    / \Phi_{0} \right] / \Phi_{0}^{2} , &
    \Phi \geq \Phi_{\rm min} \\
    0 \, , & \Phi < \Phi_{\rm min}
  \end{array}  \right.
  \label{eq:probE}
\end{equation}
where the mean flux is given by
$\langle \Phi \rangle = \Phi_{\rm min} + 2\Phi_0$.\footnote{%
The shape of this distribution is illustrated in Figure \ref{fig04}
below.}
The measurements shown in Figure 3 of \citet{Hg08} provided
constraints on the functional form of Equation (\ref{eq:probE}),
as well as values for $\Phi_{\rm min} = 2 \times 10^{18}$ Mx and
$\langle \Phi \rangle = 9 \times 10^{18}$ Mx.
These values uniquely specify the value of the exponential slope
$\Phi_{0} = 3.5 \times 10^{18}$ Mx.

In order for the code to sample from the above distribution, we
computed the cumulative probability distribution by integrating
Equation (\ref{eq:probE}) numerically.
A parameterized functional fit to the inverse of the cumulative
distribution was then found which allows a uniform random variable
(between 0 and 1) to be mapped into a proper sampling of
$P_{\rm E}(\Phi)$.
Once a random value of $\Phi$ has been chosen in this way from the
distribution, we divided the absolute flux equally between
the two poles.
We note that because the sampling from the distribution is random,
and because $N_{\rm em}$ has been truncated to be an integer, the
exact same amount of flux does not emerge in each time step.
However, over many time steps the specified emergence rate $E$ is
maintained on average.

For each emerging bipole, the $x$ and $y$ positions of the positive
pole are determined randomly.
The position of the negative pole is displaced from the positive
pole by a horizontal distance $D$ and a random orientation angle.
The separation $D$ must be large enough that the poles will not
immediately cancel one another out.
We assume that $D$ scales with the size of the flux element $r_c$,
such that $D = 1.5 r_{c} p$, where $p$ is the dimensionless
proximity factor that sets the scale for merging and cancellation
(see Section \ref{sec:3.4}).
Since $D > r_{c} p$, the poles are constrained to be noninteracting.
For this calculation we use the total flux in the entire bipole in
the definition of $r_c$ (Equation (\ref{eq:rc})), so for the mean
$\langle \Phi \rangle$, the mean separation $D$ is 6.8 Mm.
This value of $D$ is within the rather wide observational range
of separations for newly emerged ER bipoles (approximately 2--10 Mm),
as summarized by \citet{Hg01}.
Note that \citet{Hg01} found that $D \propto \Phi^{0.18}$, which
is a weaker dependence than what we assumed ($D \propto \Phi^{0.5}$)
by using Equation (\ref{eq:rc}).

\subsection{Horizontal Motions of Flux Elements}
\label{sec:3.3}

Magnetic flux concentrations are observed to move around on the
solar surface in response to plasma flows that occur on scales
ranging from narrow intergranular lanes (0.05--0.1 Mm) up to
the supergranular network ($\sim$30 Mm).
Our models were designed to test the assumption that
much of the structuring on the largest scales is a natural
by-product of smaller-scale motions \citep[see also][]{Ch07}.
Thus, the motions of flux elements are assumed to be of a diffusive
character and dominated by granule-scale (1--2 Mm) horizontal step sizes.
This stands in contrast to other Monte Carlo models of the magnetic
carpet \citep[e.g.,][]{Pn01,Si01} in which the motions of the
elements are influenced by an imposed supergranular flow pattern.

For each time step $\Delta t$, we describe the horizontal motion
of a flux element as a linear trajectory with speed $v$ and a
random orientation angle in the $x$--$y$ plane.
The orientation angle is recomputed in each time step with no
memory of its previous value, so that the long-term trajectory
of an element is essentially a ``random walk.''
Observationally, the horizontal speeds are known to depend on the
absolute fluxes in the elements, with higher-flux concentrations
tending to move with lower speeds.
Thus, we used a standard exponential fit for the mean speed $v_0$,
\begin{equation}
  v_{0} \, = \, v_{\rm weak} \, \exp \left(
  - \frac{| \Phi |}{3 \times 10^{19} \, \mbox{Mx}} \right) \,\, ,
\end{equation}
where the constant of $3 \times 10^{19}$ Mx in the denominator
is consistent with observations \citep{Hg99} and earlier models
\citep{Sj01}.
The constant $v_{\rm weak}$ is the mean speed in the limiting case of
$|\Phi| \rightarrow 0$, and it is a key free parameter in these models.
The BONES code computes the instantaneous speed $v$ for each flux
element by sampling a random number from a normal distribution
having a mean value of $v_0$ and a standard deviation of $0.3 v_{0}$
about the mean \citep[see][]{Pn01}.
When the horizontal motion is imposed on the $x$ and $y$ positions
of each flux element, the code assumes periodic boundary conditions
along the edges of the (200 Mm)$^{2}$ photospheric box.
This is designed to take account of elements that enter and leave
the box via diffusive motions.

If the horizontal motions were classically diffusive in character,
the spatial step size $\Delta r$ could be expressed as
\begin{equation}
  \Delta r \, = \, \sqrt{4 {\cal D} \, \Delta t}  \,\, ,
\end{equation}
where the diffusion coefficient ${\cal D}$ is a constant that
should not depend on the time step $\Delta t$ \citep[see][]{Sj01}.
The instantaneous velocity over a single time step would just be
$v = \Delta r / \Delta t$.
Solar observations have given rise to a large range of values for
${\cal D}$, from 50--100 km$^2$ s$^{-1}$ on granular scales to
200--2000 km$^2$ s$^{-1}$ on larger scales
\citep[e.g.,][]{Bg98,Hg99,GJ04}.
For our adopted time step of $\Delta t = 300$ s, the above range
gives values of $v$ between about 0.8 and 5 km s$^{-1}$.

On granular scales, there is evidence that the horizontal motions
do {\em not} obey classical diffusion.
\citet{Ca99} found that, for displacement times $\Delta t$ between
about 0.1 and 22 min, the mean-squared displacement $\Delta r^{2}$
does not scale linearly with $\Delta t$, but instead
\begin{equation}
  \Delta r^{2} \, \approx \, 57500 \left(
  \frac{\Delta t}{1 \, \mbox{min}} \right)^{0.76}
  \,\,\, \mbox{km}^{2} \,\, .
\end{equation}
For $\Delta t = 5$ min, this corresponds to an effective velocity
$v \approx 1.5$ km s$^{-1}$.
However, as one examines smaller displacement times, the
instantaneous velocity is larger.
For $\Delta t = 0.1$ min, $v$ increases up to 16.7 km s$^{-1}$.
The observed ``subdiffusive'' character of the horizontal motions
is believed to be related to the constraint that flux elements
must follow the narrow intergranular lanes.
Thus, it is not completely valid to model the motions as a random
walk in a two-dimensional plane that ignores the existence of
coherent granules.
In reality the elements are constrained to a fractal dimension
between 1 and 2 \citep{Ca99}.
Even the choice of a single value for $v$ may not fully reflect
the end-product of unresolved motions taking place within a time step.

In any case, it is useful to choose a representative value for the
parameter $v_{\rm weak}$ that can best reproduce the net dispersal
of granule-scale magnetic flux over many time steps.
The above analysis gives a broad range of plausible choices for
$v_{\rm weak}$ between about 0.5 and 20 km s$^{-1}$.
Several trial runs of the BONES code were produced with velocities
in this range, and a final optimized value of
$v_{\rm weak} = 6$ km s$^{-1}$ was found to produce the most
realistic solar conditions.
Section \ref{sec:4} discusses the results of models constructed
with this parameter choice.

\subsection{Merging and Cancellation}
\label{sec:3.4}

In each time step of the simulation, the horizontal distance
between every unique pair of flux elements is computed.
If the inter-element distance for a pair is less than a
prescribed critical value, we assume the flux elements coalesce
together or cancel one another out.
In a computational sense, mergings (for like polarities) and
cancellations (for opposite polarities) are treated in the same way.
The flux in the single remaining element is given by the
sum of the two signed fluxes in the original elements.
The position of this remaining element is given by the position
of the original element that had the larger absolute flux.
If an exact cancellation takes place between elements with 
equal and opposite fluxes, then both elements are assumed to
disappear from the simulation.

In order to compute the critical distance between a given pair
of elements, each element is assumed to have a ``radius of
influence'' given by $r_{c} p$, where the constant $p$ is a
dimensionless proximity factor and $r_c$ is defined in
Equation (\ref{eq:rc}).
The critical distance is the sum of the two radii of
influence for a pair of elements.

The proximity factor $p$ is another key free parameter of our
Monte Carlo simulations.
\citet{Pn01} essentially assumed that $p \approx 2.3$ based on
an empirical Gaussian profile of field strength across each flux
element.
\citet{Sj01} estimated the critical mean-free path for interactions
between average flux concentrations (in quiet network) to be about
4.2 Mm.
In order to compute a radius of influence consistent with this
mean separation (i.e., $r_{c} p = 2.1$ Mm), we can assume that the
two elements each have a mean flux
$\langle \Phi \rangle = 9 \times 10^{18}$ Mx and then use
Equation (\ref{eq:rc}) to solve for $p \approx 4.6$.
A series of trial runs of the BONES code gave rise to an optimal
value of $p=10$ that produced the most realistic solar conditions
(i.e., absolute flux densities and number distributions of flux
elements that agree with the observations discussed in Sections
4.1--4.2).
Thus, for the mean element with 
$\langle \Phi \rangle = 9 \times 10^{18}$ Mx,
its radius of influence in the models is 4.5 Mm.

The BONES code imposes lower and upper limits on the radii of
influence for the weakest and strongest flux elements, respectively.
For elements with very low fluxes, the radius of influence is not
allowed to become smaller than a typical granule size of 1 Mm.
We assume that the smallest intergranular flux tubes can easily
traverse the intergranular lanes and interact in ways that are
not resolved explicitly here \citep{Ku10}.
For the strongest flux elements, the radius of influence is not
allowed to become larger than 10 Mm.
Observationally, there do not appear to be any mergings or
cancellations that occur on spatial scales larger than this
\citep[see, e.g.,][]{Li85}.
Practically, though, the imposition of this upper limit prevents
the occurrence of ``long-range'' interactions that would be
inconsistent with the existence of the supergranular network.

Note that the actual rate of cancellation cannot be specified
explicitly in these simulations.
As described by \citet{Pn01}, the overall cancellation rate
is the eventual result of how rapidly the flux elements emerge,
move around, and interact with one another.
In a steady state, the cancellation rate eventually comes into
dynamical equilibrium with the rate of emergence $E$.
Thus, our use of the larger values of $E$ from \citet{Hg08}
implies much more rapid cancellation than was found in earlier
models such as \citet{Pn01} and \cite{Si01}.

\subsection{Spontaneous Fragmentation}
\label{sec:3.5}

Observations have shown that magnetic flux elements often split up
spontaneously into several pieces \citep[e.g.,][]{Bg96}.
Convective overturning motions on granular scales may exert stress
on the (usually intergranular) flux elements and pull them apart.
The physical processes responsible for fragmentation are not yet
understood, but magnetic reconnection may be occurring at some
stage of the process \citep{Ry03}.
There appears to be an observed relationship between the rate of
fragmentation and the total flux in an element \citep{Sj97}.
However, this applies only for relatively small concentrations
with absolute fluxes below about $10^{20}$ Mx.
Larger concentrations that give rise to pores and sunspots tend to
survive for longer times, which suggests that the fragmentation
rate saturates for $|\Phi| \gg 10^{20}$ Mx \citep{Sj01}.
In our models, we estimated the probability of fragmentation
$P_{\rm F}$ (per unit time) to be
\begin{equation}
  P_{\rm F} (\Phi) \, dt \, = \,  \frac{k_{0} |\Phi| \, dt}
  {\sqrt{1 + (|\Phi / \Phi_{\rm th}|)^{2}}}
  \label{eq:probF}
\end{equation}
where the threshold flux for saturation is given by
$\Phi_{\rm th} = 3 \times 10^{19}$ Mx.
This is a slightly simpler version of the parameterization given
by Equation (A6) of \citet{Sj01}.
The mean time between fragmentations is given by $P_{\rm F}^{-1}$.
In the limit of the largest fluxes, the mean time approaches
a constant value of $(k_{0} \Phi_{\rm th})^{-1}$.

\citet{Sj97} and \citet{Sj01} used a combination of measurements
and models to find values for $k_0$ between $4 \times 10^{-25}$
and $6 \times 10^{-25}$ Mx$^{-1}$ s$^{-1}$.
However, these were based on the same long-cadence magnetogram
observations that led to significant underestimates in the
emergence rate $E$ (see Section \ref{sec:3.2}).
Thus, we decided to increase $k_0$ by approximately the same
relative amount that $E$ was increased from the earlier values.
The models presented below all use a value of
$k_{0} = 3.5 \times 10^{-24}$ Mx$^{-1}$ s$^{-1}$.

We recompute the probabilities of fragmentation for all flux
elements in each time step of the BONES code.
For cases when a uniform-deviate random number (between 0 and 1)
is less than the probability $P_{\rm F} \Delta t$, the code
splits the flux element into two pieces.
The original element keeps a random fraction of its original
flux (constrained to be between 0.55 and 0.999), and the new
element gets the remainder of the flux.
The position of the original element stays the same, and the
new one is positioned a distance $D$ away, with a random
orientation angle.
This distance $D$ is the same value discussed in
Section \ref{sec:3.2}, and it is large enough to prevent
subsequent merging between the two new flux elements.

\section{Photospheric Field Evolution: Results}
\label{sec:4}

In this section we present results from a series of models for the
photospheric magnetic field as computed by the BONES code.
A series of tests was first performed to make sure the code was
actually evolving the flux elements as desired.
Once the tests verified that each individual process was being
modeled correctly, runs were performed that included all of the
processes together.
We created a basic set of 11 models with the main adjustable
parameter being the flux imbalance ratio $\xi$.
The input values of $B_{\rm net}$ for each of these models were
iterated until the final models had steady-state values of
$\xi$ equal to the desired input values of 0, 0.1, 0.2, 0.3, 0.4,
0.5, 0.6, 0.7, 0.8, 0.9, and 0.99
(see Equation (\ref{eq:xifit})).
Each model used a different integer as a unique seed for the
random number generator.

As described above, our final Monte Carlo models contained a much
larger emergence rate $E$ than did the earlier simulations of
\citet{Pn01} and \citet{Si01}.
If all other adjustable parameters had been kept the {\em same} as
in those models, a much larger time-steady magnetic flux would have
accumulated in the simulation box over time; i.e., the averaged
flux densities would have been much larger than the typical values
of 3--10 Mx cm$^{-2}$ observed in quiet regions and coronal holes.
In order to keep the flux density low, magnetic concentrations need
to be destroyed as rapidly as they are injected from below.
This is why the BONES code was run with more rapid horizontal
diffusion ($v_{\rm weak} = 6$ km s$^{-1}$), more sensitive merging
and cancellation ($p=10$), and more rapid fragmentation
($k_{0} = 3.5 \times 10^{-24}$ Mx$^{-1}$ s$^{-1}$) than were used
in the earlier models.
Time will tell if these parameters accurately represent the real Sun,
but as long as the emergence rate is high, the models need to
facilitate a similarly high rate of cancellation in order to produce
a realistic steady state.

Below we present results concerning the overall time-steady
photospheric magnetic fields in the simulations
(Section \ref{sec:4.1}), the statistical number distributions of
flux elements (Section \ref{sec:4.2}), and the natural production
of supergranular magnetic structures from the smaller-scale
granular motions (Section \ref{sec:4.3}).

\subsection{General Properties of the Models}
\label{sec:4.1}

The BONES models were evolved in time, using a step size of
$\Delta t = 300$ s, for a total simulation time usually exceeding
100 days and sometimes exceeding 1000 days (i.e., $10^4$--$10^5$
time steps).
Over the first 10--20 days of a simulation, sufficient magnetic
flux is injected so that the initial conditions are completely
``forgotten'' and the magnetic field reaches a state of time-steady
dynamic equilibrium.
Thus, whenever we calculate quantities that are meant to represent
the time-steady parts of a simulation (e.g., means and
standard deviations), we take only $t \geq 30$ days.
In the simulated area $A$, the total number of flux elements in
the time-steady state tends to average between 100 and 200.
Although the mean absolute flux per {\em injected} flux element
was $\langle \Phi \rangle/2 = 4.5 \times 10^{18}$ Mx, the
eventual mean flux per element in the steady state ended up
being about a factor of two larger (see below).

\begin{figure}
\epsscale{1.09}
\plotone{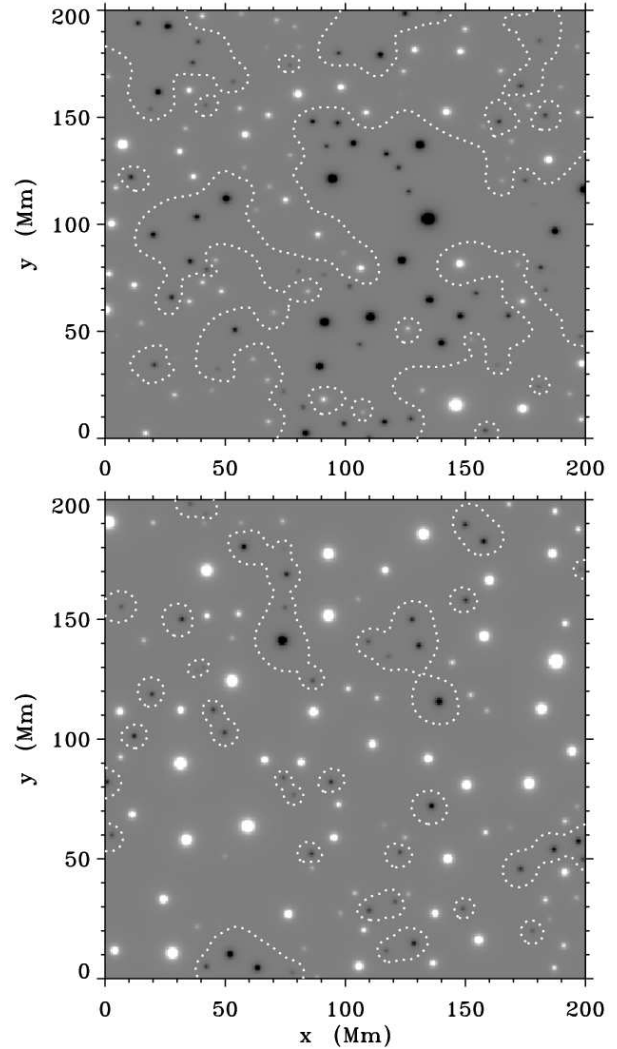}
\caption{Simulated photospheric magnetograms for random time steps
in a quiet Sun simulation with $\xi = 0$ (top) and a coronal hole
simulation with $\xi = 0.8$ (bottom).
Positive polarities are shown as white, negative polarities are
shown as black (each saturated at $|B_{z}| = 100$ G), and the
locations of magnetic neutral lines (where $|B_{z}| = 0$) are
overplotted as white dotted curves.}
\label{fig01}
\end{figure}
Figure \ref{fig01} shows simulated magnetogram images for
representative time snapshots in two of the models: one for
a region of balanced magnetic flux ($\xi=0$) and one for a
large degree of imbalance ($\xi=0.8$).
The continuous magnetic field strength at the photosphere ($z=0$)
was calculated using the multiple monopole model described in
Section \ref{sec:5.1}.
A medium gray shade denotes $B_{z} \approx 0$, and the saturation
to white and black is imposed at $B_{z} = +100$ and $-100$ G,
respectively.
For the balanced case, the neutral line meanders through the domain
stochastically and splits the region into two roughly equal areas.
For the imbalanced case, the neutral lines surround and confine the
regions of minority polarity.

The balanced ``quiet Sun'' model shown in Figure \ref{fig01}(a)
has an average total number of flux elements $N = 163$, with
roughly equal numbers of positive and negative elements and an
average absolute flux per element of $8.9 \times 10^{18}$ Mx.
The imbalanced ``coronal hole'' model shown in Figure \ref{fig01}(b)
has an average total $N = 122$, with approximately 81 of the
elements being positive and 41 being negative.
Note that if the absolute flux per element was equal for the
positive and negative populations, we would have expected that
$N(1 + \xi)/2 = 110$ elements would be positive, and
$N(1 - \xi)/2 = 12$ elements would be negative.
Since the number of positive [negative] elements is smaller [larger]
than predicted, it is clear that the two populations must have
different average absolute fluxes.
In fact, for the $\xi=0.8$ model, the average fluxes per element in
the positive and negative sets were $2.3 \times 10^{19}$ Mx and
$5.1 \times 10^{18}$ Mx, respectively.

\begin{figure}
\epsscale{0.95}
\plotone{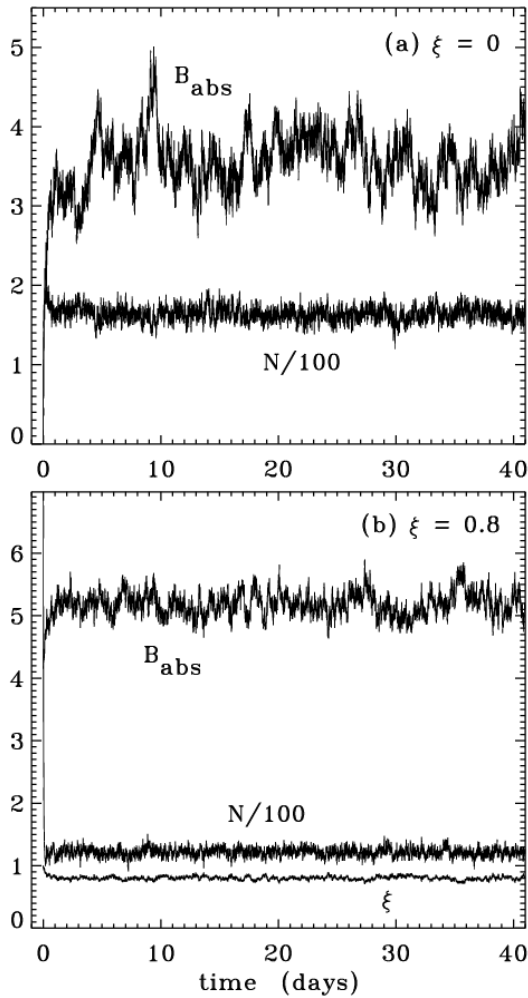}
\caption{Time evolution of statistical quantities in the
(a) $\xi=0$ and (b) $\xi=0.8$ photospheric models.
The temporal variability of the box-averaged
absolute flux density $B_{\rm abs}$, the total number $N$
of flux elements in the simulation (divided by 100 to keep the
curve in the same plotting domain as the other curves), and the
flux imbalance ratio $\xi$ are shown as labeled.}
\label{fig02}
\end{figure}
In Figure \ref{fig02} we plot the time dependence of several
statistical quantities for the $\xi=0$ and $\xi=0.8$ cases.
These models reached dynamical equilibrium in only about 5 days of
simulation time, and only the first 40 days are shown.\footnote{%
By ``dynamical equilibrium'' we mean that there appears to be
a time-steady mean state existing together with substantial
variations about that mean.
It also seems clear that no single ingredient in the
photospheric flux evolution model is responsible for determining
these time-steady mean properties.
This state is a complex, nonlinear {\em dynamic balance} between
emergence, merging, cancellation, diffusion, and fragmentation.}
After a stochastic steady state has been established, the level of
continuing temporal variability appears similar in character to the
simulations of \citet{Pn01} and \citet{Ch07}.
Note that the imbalance ratio $\xi$ does not approach a rigidly
constant value, but instead fluctuates with a standard deviation
that is typically 2\%--10\% of its mean value.

Comparing Figures \ref{fig02}(a) and \ref{fig02}(b), we see that
as $\xi$ increases the mean of the absolute flux density
$\langle B_{\rm abs} \rangle$ increases and its variance decreases.
Larger values of $\xi$ correspond to lower rates of flux emergence
(see Equation (\ref{eq:Efit})), so that a typical flux element in
the large-$\xi$ simulation tends to have a longer lifetime before
it is destroyed.
However, the functional form of $E(\xi)$ is not the only reason
for the increase in $B_{\rm abs}$ with increasing $\xi$.
It is possible to illustrate such an increase with a simple analytic
model that assumes a {\em constant} emergence rate.
If the emergence rate $E$ is fixed, but the box-averaged rate of
cancellation is assumed to be proportional to the product of the
positive and negative flux densities present in the box, then their
time evolution can be approximated to be a simple balance between
these two effects, with
\begin{equation}
  \frac{\partial B_{+}}{\partial t} \, = \,
  \frac{\partial |B_{-}|}{\partial t} \, = \,
  E - C B_{+} |B_{-}|  \,\, .
\end{equation}
In a steady state, the time derivatives can be ignored and we can
solve for $E = C B_{+} |B_{-}|$.
The individual values of the constants $E$ and $C$ do not need to
be specified explicitly, but let us assume their ratio $E/C$ is
a known constant called $B_{0}^{2}$.
Thus, it becomes possible to solve for the absolute flux density
in closed form,
\begin{equation}
  B_{\rm abs} \, = \, B_{+} + |B_{-}| \, = \,
  \frac{2 B_{0}}{\sqrt{1 - \xi^{2}}}  \,\, .
  \label{eq:Banalytic}
\end{equation}
The above expression shows how $B_{\rm abs}$ must increase with an
increasing imbalance ratio $\xi$, even in the case where $E$ is
independent of $\xi$.

\begin{figure}
\epsscale{1.09}
\plotone{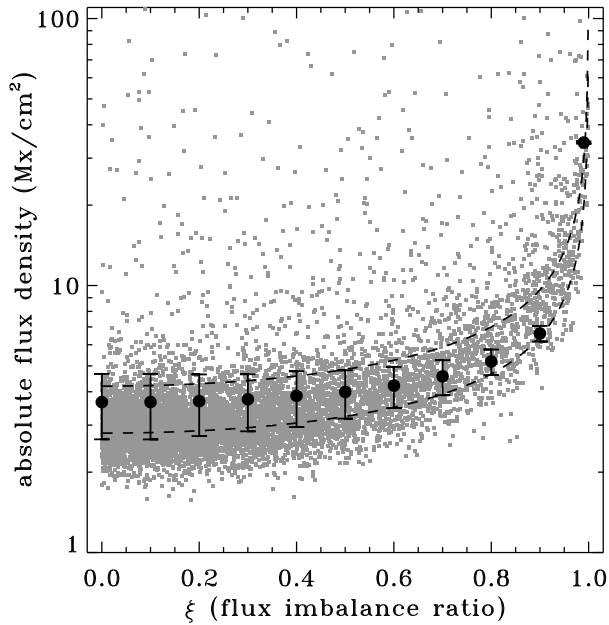}
\caption{Steady-state dependence of absolute flux density
$B_{\rm abs}$ on the flux imbalance ratio $\xi$.
Mean results from the BONES simulations (filled circles, with
$\pm 3 \sigma$ error bars) are compared with observational data
computed from SOLIS full-disk magnetograms (gray points), and with
the analytic approximation given by Equation (\ref{eq:Banalytic})
(dashed curves).}
\label{fig03}
\end{figure}

Figure \ref{fig03} shows how the time-steady values of
$\langle B_{\rm abs} \rangle$ from the simulations vary as a
function of $\xi$.
The error bars on these model points show $\pm 3$ standard
deviations around the mean values.
To ensure that specific realizations of the random number sequences
did not affect the results, the means and standard deviations for
each value of $\xi$ were computed from three independent runs of
the BONES code.
Each run used a different random seed, and each run was performed
for a total of 400 days of simulation time.
The modeled absolute flux densities generally fall between the
observationally expected limiting values of about 3 and 10
Mx cm$^{-2}$.
Figure \ref{fig03} also shows two curves that illustrate the
functional dependence of the simple analytic estimate of
Equation (\ref{eq:Banalytic}) above.
The two curves, which were computed using the arbitrary normalization
constants $B_{0} = 1.4$ and 2.1 G, appear to bracket the modeled
points surprisingly well.

In Figure \ref{fig03} we also plotted measurements made by the
Vector SpectroMagnetograph (VSM) instrument of the Synoptic Optical
Long-term Investigations of the Sun (SOLIS) facility \citep{Kc03}.
We used publicly available full-disk longitudinal magnetograms
taken in the \ion{Fe}{1} 6301.5 {\AA} line.
Over the time period from August 2003 to November 2009, we obtained
one magnetogram per month for a total of 73 individual full-disk maps.
For each magnetogram we generated a grid of ``macropixels''
covering the central part of the solar disk (out to 0.7 $R_{\odot}$
from disk-center).
Each macropixel was defined to be $100 \times 100$ magnetogram pixels,
or 113$''$ square \citep[see also][]{Hg08}.
For each macropixel, we measured the average flux densities of the
positive and negative polarities, $B_{+}$ and $B_{-}$, and computed
$B_{\rm abs}$ and $\xi$ as defined in Section \ref{sec:3.1}.
A total of 8264 individual measured data points are shown in
Figure \ref{fig03}.

The bulk of the low field-strength SOLIS data shown in
Figure \ref{fig03} appear to follow the same general increasing
trend with $\xi$ as do the modeled points and analytic curves.
The ``long tail'' in the data points that extends upward to
10--100 Mx cm$^{-2}$ represents times when the macropixels covered
parts of active regions.
Points on the upper-left of the plot represent active regions that
were mostly centered in the macropixel, and points on the upper-right
represent times when only one dominant polarity of an active region
was in the macropixel.
The models presented in this paper are generally meant to be
simulations of quiet Sun and coronal hole regions, which are sampled
by the majority of weak-field data points in the lower part
of Figure \ref{fig03}.

\subsection{Number Distributions of Flux Elements}
\label{sec:4.2}

An additional way to verify that the BONES simulations produce
magnetic fields similar to those on the real Sun is to examine
the {\em probability distributions} of element fluxes and compare
them with observed distributions.
Because the simulations typically have only 100--200 elements in
them at any one time, we sampled the distributions a number of times
in order to accumulate statistics appropriate for a large number of
uncorrelated patches of Sun.
In the models, the time cadence for this sampling was fixed at 30 days.
This time cadence was found to be more than adequate for the
requirement that any given distribution of flux elements must be
completely recycled from (i.e., uncorrelated with) the distribution
at the previous sampling time.
For each case discussed below, the simulations were run until the
total number of collected flux elements exceeded $10^5$.

\begin{figure}
\epsscale{1.00}
\plotone{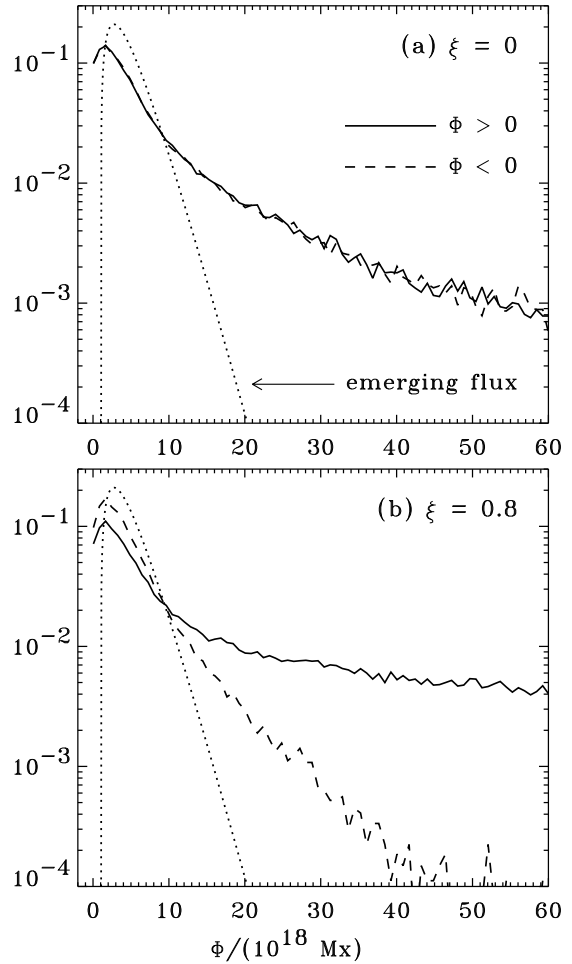}
\caption{Statistical number distributions of flux elements as
a function of their absolute fluxes in the
(a) $\xi=0$ and (b) $\xi=0.8$ models.
The time-steady distributions in the numerical simulations are
shown separately for positive (solid curves) and negative
(dashed curves) polarities, and both are compared with the
imposed distribution of emerging flux elements (dotted curves).
For plotting convenience, both the fluxes themselves and the
normalized probability distributions were divided by $10^{18}$.}
\label{fig04}
\end{figure}

Figure \ref{fig04} shows example distributions for the two models
discussed above ($\xi = 0$ and 0.8).
The distributions of positive and negative polarity elements are
plotted separately.
For comparison, the analytic distribution of emerging flux elements
given by Equation (\ref{eq:probE}) is also shown.
This latter distribution has been scaled down in flux by a factor of
two (i.e., shifted to the left in the plot) to show the distribution
of fluxes in the individual poles of the emerging bipoles, not the
total absolute flux in the bipoles as specified by
Equation (\ref{eq:probE}).
For ease of comparison with observations, these plots are shown
in the same general format as Figures 4 and 6 of \citet{Pn02}
and Figures 2 and 3 of \citet{Hg08}.

The time-steady distributions shown in Figure \ref{fig04} are
substantially ``flatter'' than the initial distribution of
emerging flux elements.
In other words, the fluxes have spread out from the relatively
narrow range of injected fluxes (roughly $10^{18}$--$10^{19}$ Mx)
to both lower and higher values \citep[see][]{Pn02}.
Most noticeably, the populations of flux elements with
$|\Phi| \gtrsim 3 \times 10^{19}$ Mx are hugely enhanced with
respect to the distribution of injected flux elements.
These stronger flux elements must be the result of mergings
between smaller elements of like polarity.
In addition, the existence of this enhanced strong-flux tail is
the reason that the mean flux per element is larger than the
mean flux in a newly emerged flux element (see
Section \ref{sec:4.1}).

Although it is difficult to see in the plots, there is also a
significant number of elements in the simulations with fluxes
{\em below} the minimum emergent flux per element
($\Phi_{\rm min}/2 = 10^{18}$ Mx).
These weakest flux elements must be the result of fragmentation and
partial cancellation.
For the $\xi=0$ case, 22\% of the flux elements have fluxes less
than this threshold value.
Because of their small fluxes, however, these account for only
about 2.7\% of the total absolute flux in the simulation.
For the $\xi=0.8$ case, 18\% of the flux elements have fluxes below
the emerging threshold value, and they account for 1.2\% of the
total absolute flux.

Figure \ref{fig04}(b) shows the difference between the distributions
of positive and negative elements for the imbalanced case of
$\xi = 0.8$.
Overall, the majority polarity has a flatter distribution than
does the minority polarity, but there is an excess of minority
polarity elements for the weakest fluxes
($|\Phi| \lesssim 10^{19}$ Mx).
This is in good agreement with the observational conclusions of
\citet{Zh06} for coronal holes.
Also, the differences in shape shown in Figure \ref{fig04}(b) are
highly reminiscent of the flux element distributions shown in
Figure 2 of \citet{Hg08} for coronal holes.

\subsection{Naturally Occurring Supergranular Scales}
\label{sec:4.3}

The resemblance between the cellular pattern of solar granulation
and that of the larger-scale supergranulation has long been
interpreted as evidence that both phenomena are manifestations of
the Sun's convective instability \citep[e.g.,][]{Le62,RT79,SW91,RR10}.
However, because the flow patterns in the supergranular network
are weak and intermittent, it has not been possible to definitively
prove their convective origin.
It may be that multiple interactions between granule-scale
structures produce a distributed network of downflows that in turn
seeds horizontal supergranular flows and the aggregation of strong
network fields \citep{Ra03,Gb09}.
Alternately, the opposite may be the case; i.e., it may be the
aggregation of small-scale magnetic fields that gives rise to the
weak supergranular flows \citep{Ch07}.
In this section we show that the BONES simulations provide some
evidence for the initial magnetic-field aggregation described in
the latter scenario.

How are the spatial scales of supergranulation measured?
It is well known that the dominant cell sizes are of order 10--30 Mm,
but different types of measurement give different answers.
\citet{SL64} found cell diameters around 32 Mm by interpreting
autocorrelation functions of chromospheric Dopplergrams.
\citet{SB81} traced the cells manually, based on \ion{Ca}{2}
K-line intensity images, and found diameters of $\sim$22 Mm.
\citet{Wh88} and \citet{Wh96} applied the autocorrelation
technique to magnetograms and found scale sizes between 10 and 25 Mm,
depending on the precise diagnostic techniques used.
Finally, \citet{DT04} and \citet{Hg97} used a range of sophisticated
algorithms to trace and characterize supergranular boundaries, and
found average diameters of only $\sim$15 Mm.

\begin{figure}
\epsscale{1.09}
\plotone{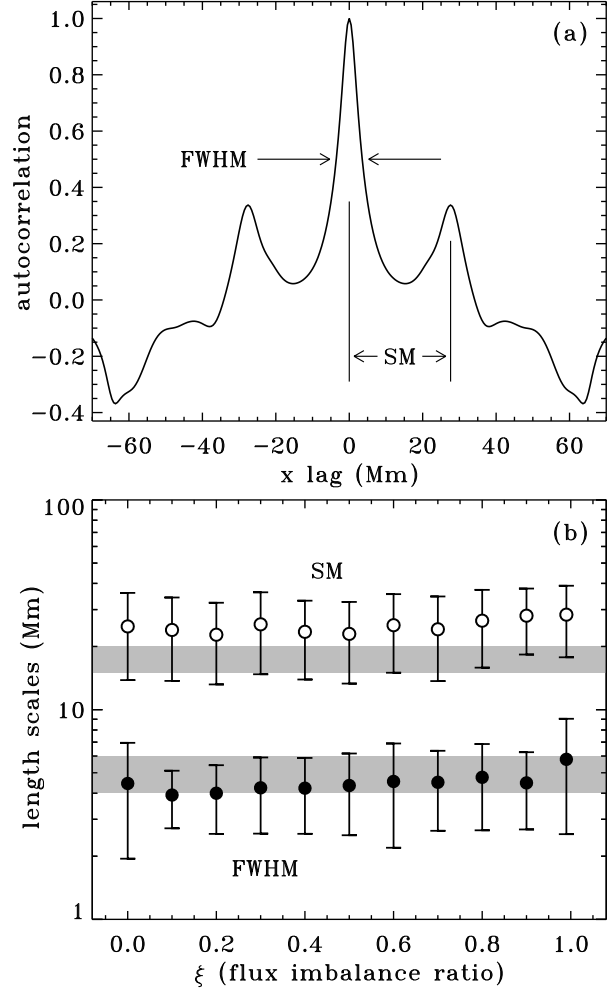}
\caption{(a) Example of a simulated magnetogram autocorrelation
function for a slice across the $\xi=0$ model, plotted as
a function of the lag parameter $x'$ (see Equation (\ref{eq:ac})).
(b) Results for modeled mean values of FWHM (filled circles)
and SM (open circles) plotted as a function of $\xi$, with
error bars denoting $\pm 1 \sigma$ in the simulated distributions
of values, and the observed ranges of FWHM and SM values from
\citet{Wh88} (gray regions).}
\label{fig05}
\end{figure}

Because the BONES simulations predict only the properties of the
magnetic field---and neither the chromospheric emission nor the
Doppler velocities---we decided that the most straightforward
comparison to make would be with the measured magnetogram
autocorrelation functions of \citet{Wh88}.
First, a random time step from each of the 11 models was used
to create simulated magnetograms similar to those shown in
Figure \ref{fig01}.
Then, for each $y$ row in the magnetogram, we computed a series of
one-dimensional autocorrelation functions in the $x$ direction for
the scalar value of $B_{z}$, i.e.,
\begin{equation}
  AC (x' , y) \, = \, \int_{-\infty}^{+\infty}
  B_{z} (x,y) \, B_{z}(x + x' , y) \, dx \,\, ,
  \label{eq:ac}
\end{equation}
which was then normalized such that $AC(0,y) = 1$.
Figure \ref{fig05}(a) shows an example autocorrelation function
from the $\xi=0$ simulation, plotted as a function of the lag
parameter $x'$.
Similar results were found when the roles of the $x$ and $y$
coordinates were reversed.

We characterized the model autocorrelation functions by finding both
the full-width at half-maximum (FWHM) of the central peak and the
distance between the central peak and the next secondary maximum (SM).
Doing this for each value of $y$ gave rise to ensembles of values
for FWHM and SM in each of the 11 simulations.
Figure \ref{fig05}(b) shows the mean values for each of these
ensembles, along with error bars that show $\pm 1$ standard
deviations about the means.
There is no significant $\xi$ dependence in the modeled values.
For all 11 simulations, the average model FWHM is 4.48 Mm and the
average SM distance is 25.1 Mm.
These values compare favorably to the solar observations reported by
\citet{Wh88} (shown as gray bars in Figure \ref{fig05}), who found
FWHM values between 4 and 6 Mm, and SM distances of 15 to 20 Mm.

The benefit of making a direct comparison between simulated and
observed FWHM and SM values is that there is no need to interpret
these quantities in terms of arbitrarily defined cell
diameters.\footnote{%
See, however, Figure \ref{fig08} below for a more intuitive way
of visualizing the naturally occurring ``supergranular network''
in these simulations.}
The models appear to succeed in roughly reproducing the
observed autocorrelation properties of the network.
It may be possible to explain this success by invoking processes
of {\em diffusion-limited aggregation} as suggested by \citet{Ch07}.
In this picture, time-steady magnetic structures ``collect'' on
specific scales that depend on the combined emergence, diffusion,
and cancellation of flux elements.
Supergranular flows may then occur as a result of the magnetic
structuring.
\citet{Ch07} performed tests with a Monte Carlo model that varied
several of the discrete step sizes and interaction distances, and
found that the resulting supergranular scale size does not depend
on these input parameter choices.
Instead, it is the overall level of flux emergence and horizontal
diffusion---which in turn drives the cancellation rate---that
sets the time-steady distance between network concentrations.

\section{Coronal Field Evolution}
\label{sec:5}

One of the major goals of this paper is to explore how the complex
photospheric fields in the magnetic carpet connect with time-variable
open flux tubes and closed loops in the extended corona.
Thus, here we describe how the field lines are traced upwards and
are evolved in time (Section \ref{sec:5.1}),
we summarize the resulting open and closed fields as a function of
the flux imbalance ratio $\xi$ (Section \ref{sec:5.2}),
we compute relevant time scales for the opening up of closed
flux tubes (Section \ref{sec:5.3}), we estimate the amount of
magnetic energy that emerges in the form of bipoles (Section
\ref{sec:5.4}), and we compare it to the energy released into the
solar wind by magnetic reconnection (Section \ref{sec:5.5}).

\subsection{Field-line Extrapolation Method}
\label{sec:5.1}

As summarized in Section \ref{sec:2}, we compute the vector magnetic
field ${\bf{B}}$ above the photospheric surface by assuming the
field is derivable from a scalar potential.
In other words, each flux element is assumed to act as a monopole-type
source, with
\begin{equation}
  {\bf B} ({\bf r}) \, = \, \sum_{i} \frac{\Phi_{i}}{2\pi}
  \frac{{\bf r} - {\bf r}_{i}}{| {\bf r} - {\bf r}_{i} |^3} \,\, ,
  \label{eq:Bdef}
\end{equation}
where the coordinates ${\bf r}_{i} = (x_{i}, y_{i}, z_{i})$
specify the locations of each flux element $i$, and the field
point ${\bf r} = (x,y,z)$ can be located anywhere at or above
the photosphere ($z \geq 0$).
$\Phi_i$ is the signed magnetic flux in each element
\citep[see, e.g.,][]{Wa98,Cl03}.

To avoid singularities at the solar surface, all elements are
assumed to be ``submerged'' below the photosphere \citep{Se86,Lg05}.
For simplicity we assumed that all flux elements are at a constant
depth.
We chose an optimum value of $z_{i} = -1$ Mm on the basis of the
following considerations.
The peak magnetic field strength $B_{\rm peak}$ in the photosphere,
due to a single flux element, occurs right over the point itself at
$x = x_i$, $y = y_i$, and $z = 0$.  Thus,
\begin{equation}
  B_{\rm peak} \, = \, \frac{\Phi_i}{2\pi z_{i}^2}  \,\, .
  \label{eq:Bpeak}
\end{equation}
We want to ensure that $|B_{\rm peak}|$ is less than the equipartition
field strength $B_{\rm max}$ for all elements in the simulation
(see Section \ref{sec:3.1}).
Because we do not model pores and sunspots, we can apply this
constraint to elements up to a maximum flux of
$|\Phi| \approx 10^{20}$ Mx.
Thus, applying the condition $|B_{\rm peak}| \leq B_{\rm max}$ to
Equation (\ref{eq:Bpeak}) for this value of the flux gives rise to
$|z_{i}| \gtrsim 1.1$ Mm.
On the other hand, observations have shown that the field strength
in a recently emerged ER is at least a few hundred Gauss \citep{Ma88}.
For the average flux in one pole of an emerging ER (i.e.,
$\langle \Phi \rangle /2 \approx 4.5 \times 10^{18}$ Mx), we
apply the condition $B_{\rm peak} \gtrsim 100$ G and obtain
an upper limit $|z_{i}| \lesssim 0.85$ Mm.
The two above constraints on the magnitude of $z_i$ are formally
incompatible with one another, but the value $\sim$1 Mm appears
to be a likely compromise between the two.

The BONES code contains a subroutine that can either trace field
lines up from the photospheric surface or down from a larger height.
The incremental path length $\Delta s$ for numerical steps taken
along the field varies with height, from a minimum value of 0.03 Mm
at the photosphere to a maximum value of 10 Mm at a height of
$z=200$ Mm.  At intermediate heights,
\begin{equation}
  \Delta s \, = \, (0.03 \,\, \mbox{Mm})^{1-\zeta}
  (10 \,\, \mbox{Mm})^{\zeta} \,\, ,
\end{equation}
where $\zeta = z / (200 \,\, \mbox{Mm})$.
Field lines that begin at the photosphere are traced until they
either curve back down to intersect the $z=0$ plane again
(and are called ``closed'') or they climb past a maximum height
of 200 Mm (and are called ``open'').
As discussed in Section \ref{sec:2}, on the real Sun it is possible
that many flux tubes that reach higher than 200 Mm may eventually
be closed back down in the form of large-scale helmet streamers.
Whether this occurs or not depends on the global distribution of
magnetic flux across the entire solar surface.
In any case, it is likely that some plasma that reaches large heights
in streamers also interacts with the accelerating solar wind
\citep{Wa00}, so it may not be too erroneous to classify these field
lines as open.

When the Monte Carlo simulation of the photospheric field settles
into a dynamical steady state (defined here as $t \geq 50$ days),
we begin tracing field lines in order to compute the coronal
vector field in each time step.
This essentially assumes that any temporal changes occur
``instantaneously;'' i.e., with a time scale shorter than
$\Delta t = 5$ min.
In similar kinds of potential-field simulations, \citet{Rg09} found
that the actual delay between a given photospheric impulse and the
response higher up in the corona is only of order 2 min.
Thus, our assumption that ${\bf B}({\bf r})$ can be recomputed
from each time step's new lower boundary condition appears to be
reasonable.

In order to quantify the changes that occur in the magnetic field
from one time step to the next, we trace a set of field lines
that is associated with the $N$ flux elements on the surface.
The general idea is to compare the open/closed topology of flux
tubes that can be identified unambiguously both at the beginning
of a time step and at the end \citep[see also][]{Cl05}.
If a flux element moves around on the surface and does not undergo
substantial merging, cancellation, or fragmentation, then we can
say that it has ``survived'' that time step, and thus it makes sense
to evaluate how its open/closed connectivity may have changed.
In cases where the merging, cancellation, or fragmentation makes
only a minor change to an original element's flux, we also
consider that element to have survived when the element's flux
changes by less than a specified fractional threshold $\delta$.
In most runs of the BONES code presented below, $\delta = 0.1$.
This means that if a flux element ends the time step with a flux
that is within 10\% of its original flux, it is classified as
being the same element.
Flux elements that cannot be tagged in this way are not counted.
We discuss the effects of varying the $\delta$ parameter below.

Rather than just trace one field line from each flux element, we
instead chose to more finely resolve the coronal magnetic field
by tracing seven field lines from each element.
The initial footpoints of these seven field lines are arranged in a
hexagonal pattern with respect to each flux element's circular
``patch'' on the surface.
One field line is centered on the flux element.
The other six are arranged in a ring around the central point with
an angular separation of {60\arcdeg}, each at a horizontal
distance of $r_{c} (1+p)/2$ from the central point.
This distance is halfway between the flux element's intrinsic
radius $r_c$ and its critical interaction distance as defined
in Section \ref{sec:3.4}.
At the beginning of each time step the BONES code traces $7N$
field lines and tags each footpoint with a unique (nonzero)
numerical identifier.
Each of the flux tubes associated with element $i$ is assigned
an equal magnetic flux $\Phi_{i}/7$.
During the progress of each time step, new flux elements that
emerge are given an identifier of zero.
Also, if merging, cancellation, or fragmentation changes the
flux in an element to a degree greater than the relative
threshold $\delta$, its numerical identifier is reset to zero.
At the end of each time step, the coronal field is traced again
for the subset of surviving flux elements that have nonzero
numerical identifiers.
The magnetic flux in those elements is grouped into four bins
that are defined by whether the flux tubes were open or closed
at the beginning of the time step, and whether they are open or
closed at the end.
Section \ref{sec:5.2} discusses the distributions of magnetic flux
in those four bins.

We note that our method of accounting for the open and closed
magnetic flux has several potential shortcomings.
By not counting either the newly emerged flux elements or those
that undergo substantial merging, cancellation, or fragmentation,
we run the risk of not seeing fields that may be releasing lots of
energy via magnetic reconnection.
We will see below, though, that the magnetic-carpet evolution is
not so vigorous that these flux elements represent a significant
fraction of the total number.
In fact, for most models the fraction of magnetic flux that is
missed by not counting these ``rapid evolvers'' is only of order
5\% to 15\%.
Another possible limitation of our method is that we trace the
identities of individual flux tubes for only one time step.
If we wanted to measure more accurate time scales for flux
reconfiguration, it may have been advantageous to follow field
lines for {\em more} than just one time step.
However, since the magnetic carpet keeps evolving, the number of
flux tubes that would become uncountable (i.e., missed by virtue
of exceeding the threshold $\delta$) increases for each additional
time step over which flux-tube survival would be traced.
Following field lines only over the course of one time step, with
$\Delta t = 5$ min, gave the best balance of time resolution and
flux capturing.

\subsection{General Results}
\label{sec:5.2}

\begin{figure}
\epsscale{0.97}
\plotone{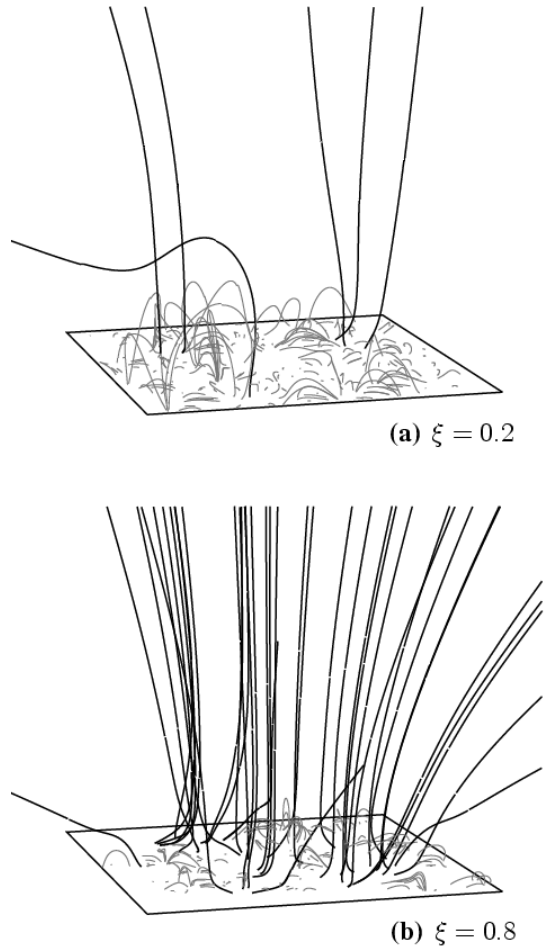}
\caption{Traced magnetic field lines at example time steps in
BONES models having (a) $\xi = 0.2$, and (b) $\xi = 0.8$.
Open and closed field lines are plotted in black and gray,
respectively.
In both panels, the horizontal box outlines the (200 Mm)$^{2}$
photospheric simulation domain.  The vertical scaling has been
stretched by about a factor of two, such that the uppermost tips
of the field lines are at a height of $z \approx 110$ Mm.}
\label{fig06}
\end{figure}
Figure \ref{fig06} illustrates a selection of field lines for
BONES models with a mostly balanced lower boundary ($\xi = 0.2$)
and a highly imbalanced lower boundary ($\xi = 0.8$).
The three-dimensional field lines are shown projected into a
two-dimensional plane that is defined by an observer viewing the
scene at an inclination angle {82\arcdeg} from the normal to the
photosphere.
Two different shades denote closed versus open field lines.
Models with more imbalanced fields (i.e., higher values
of $\xi$) have both a larger fraction of open flux and a smaller
vertical extent for the closed loops.
Both of these trends are examined quantitatively below.

We studied the statistical properties of the closed loops in the
simulations by tracing large numbers of field lines from random
starting locations $(x,y,0)$ in the photosphere.
Example time snapshots from the 11 models (with varying $\xi$ values)
were used to trace at least 5000 loops in each model.
For the six models with $\xi \leq 0.5$, for which there were fewer
open field lines, we were able to compute at least 20000 loops.
The maximum heights of these loops were collected into
11 statistical distributions, one for each model.
Although the means and standard deviations of these distributions
were computed, the distributions were far from Gaussian in shape.
Thus, we quantified them further by computing percentile intervals
$H_n$ of the sorted cumulative distributions of heights.
For example, 25\% of the loops have heights less than the quartile
height of $H_{25}$, and 50\% of the loops have heights less than the
median height of $H_{50}$.
We also computed $H_{75}$ and $H_{95}$, with the latter being an
approximate indicator of the largest loops (without being dependent
on the statistically insignificant tail of the {\em very} largest
loops).

\begin{figure}
\epsscale{1.09}
\plotone{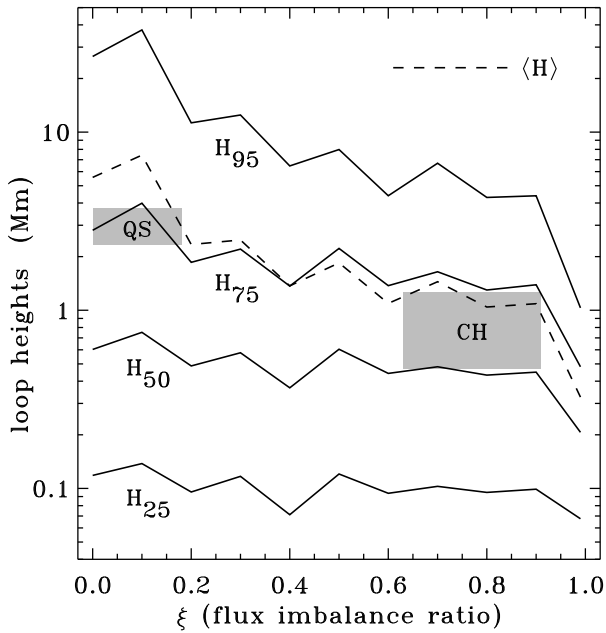}
\caption{Variation of percentile intervals of the sorted statistical
distributions of loop heights, shown as a function of $\xi$.
Percentiles at the 25\%, 50\%, 75\%, and 95\% levels (solid
curves) are compared with the mean loop height $\langle H \rangle$
(dashed curve) and with observationally inferred values from
\citet{WS04} for quiet Sun (QS) and coronal hole (CH) regions
(gray boxes).}
\label{fig07}
\end{figure}
Figure \ref{fig07} shows how the percentile intervals vary as a
function of the flux imbalance ratio $\xi$.
On the smallest spatial scales (i.e., for granule-sized loops
characterized by $H_{25}$ and $H_{50}$) there does not appear to
be a significant dependence on $\xi$.
However, the longest loops follow the trend that is visually
apparent in Figure \ref{fig06}; i.e., the more balanced the
photospheric field, the larger the loops.
This trend is apparent not only in $H_{75}$ and $H_{95}$, but also
in the mean height $\langle H \rangle$ that is weighted more
strongly by the longest loops.

Figure \ref{fig07} also shows approximate observational ranges of
mean loop heights for quiet Sun (QS) and coronal hole (CH) regions
as determined by \citet{WS04}.
These loop-height calculations were similar to ours in that they
were based on potential-field extrapolations from photospheric
lower boundary conditions, but \citet{WS04} used observed magnetograms
from the Michelson Doppler Imager (MDI) instrument on {\em SOHO}
\citep[see also][]{Cl03,Ti10,It10}.
The overall agreement with the modeled $\xi$ dependence of
$\langle H \rangle$ is good.
The general trend for high-$\xi$ CH regions to have shorter
loops than low-$\xi$ QS regions is also consistent with the
trend pointed out by \citet{Fe99} and \citet{Gl03} for the source
regions of fast solar wind to be correlated with short loops and
the source regions of slow wind to be correlated with long loops.

\begin{figure}
\epsscale{1.15}
\plotone{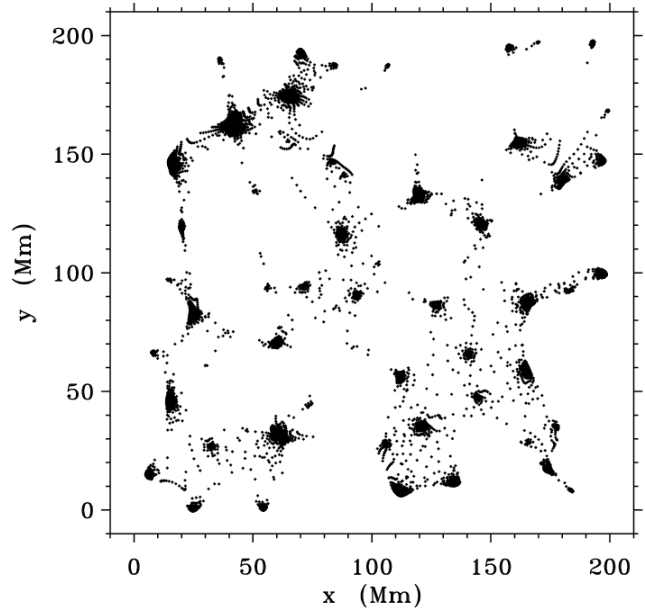}
\caption{Photospheric locations of footpoints of ``open'' magnetic
field lines traced down from an evenly spaced grid at a height of
$z = 200$ Mm, for one time snapshot of the $\xi = 0.8$ model.}
\label{fig08}
\end{figure}

A representative illustration of the footpoints of open field lines
is given in Figure \ref{fig08} for the $\xi = 0.8$ model.
This plot shows the locations of the photospheric footpoints of
$10^4$ field lines that were traced down from an evenly spaced
grid at the top ($z = 200$ Mm).
In order to account for the horizontal flaring of potential field
lines from the finite-sized simulation box, the grid of
$100 \times 100$ starting points had an overall horizontal size
of $1800 \times 1800$ Mm in the $x$ and $y$ directions
(centered on the $200 \times 200$ Mm simulation box).
The overall appearance of Figure \ref{fig08} is highly reminiscent
of the observed supergranular network.
The apparent ``cell diameters'' tend to be between 20 and 40 Mm
as on the real Sun.
Note also the appearance of thin channels, stretched between
smaller knots of closed-field regions, that appear to support the
connectivity theorems described by \citet{An07}.

All of the $10^4$ open field lines with footpoints shown in
Figure \ref{fig08} are of positive polarity.
This is the dominant polarity as specified by the initial conditions
of the BONES code (see Section \ref{sec:3.1}).
All negative polarities end up connected to positive polarities
in closed loops, and thus there are no ``open funnels'' with the
non-dominant polarity.
Of course, this is also a highly simplified situation when compared
to the real Sun, for which there are often network concentrations of
both polarities even in strongly unipolar coronal holes.

\begin{figure}
\epsscale{1.09}
\plotone{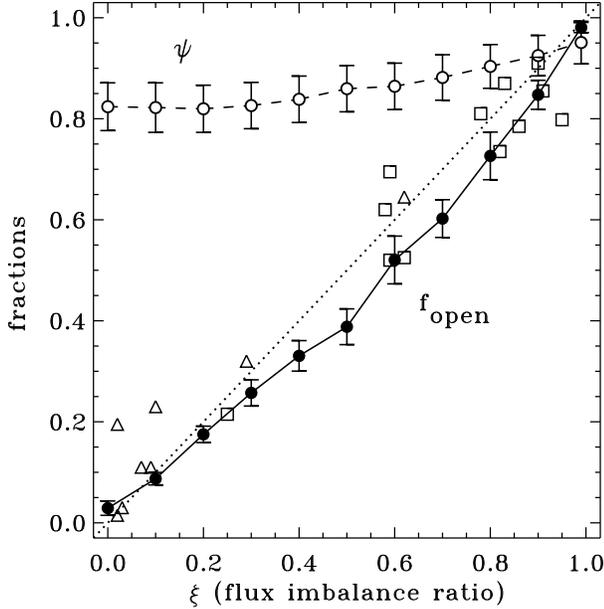}
\caption{Various dimensionless flux fractions shown as a function of
$\xi$: mean values of $f_{\rm open}$ (filled circles) and mean values
of $\psi$ (open circles), both with their $\pm 1 \sigma$ spreads shown
as error bars, and observational estimates of $f_{\rm open}$ from
\citet{WS04} in QS (triangles) and CH (squares) regions.}
\label{fig09}
\end{figure}

As described above, at the beginning of each time step there is a
set of field lines traced from each of the flux elements.
These $7N$ field lines are used to estimate the instantaneous
fractions of absolute unsigned flux that are either open or closed.
The fraction of flux that is open is denoted $f_{\rm open}$, and in
Figure \ref{fig09} we show its mean value as a function of
the $\xi$ imbalance ratio.
This fraction is never exactly the same from one time step to the
next, and the error bars show $\pm 1$ standard deviations about the
mean values.
On average, $f_{\rm open}$ is roughly equal to $\xi$ itself.
In other words, models with balanced fields tend not to have much
open flux, but when there is an increase in the unbalanced component
of the field there is a corresponding increase in the fraction of
open flux.
Figure \ref{fig09} also compares the modeled values of $f_{\rm open}$
with observational determinations of this quantity from \citet{WS04},
and there is a similar trend of direct proportionality, with
$f_{\rm open} \approx \xi$.

\subsection{Comparison of Relevant Time Scales}
\label{sec:5.3}

\begin{figure*}
\epsscale{0.93}
\plotone{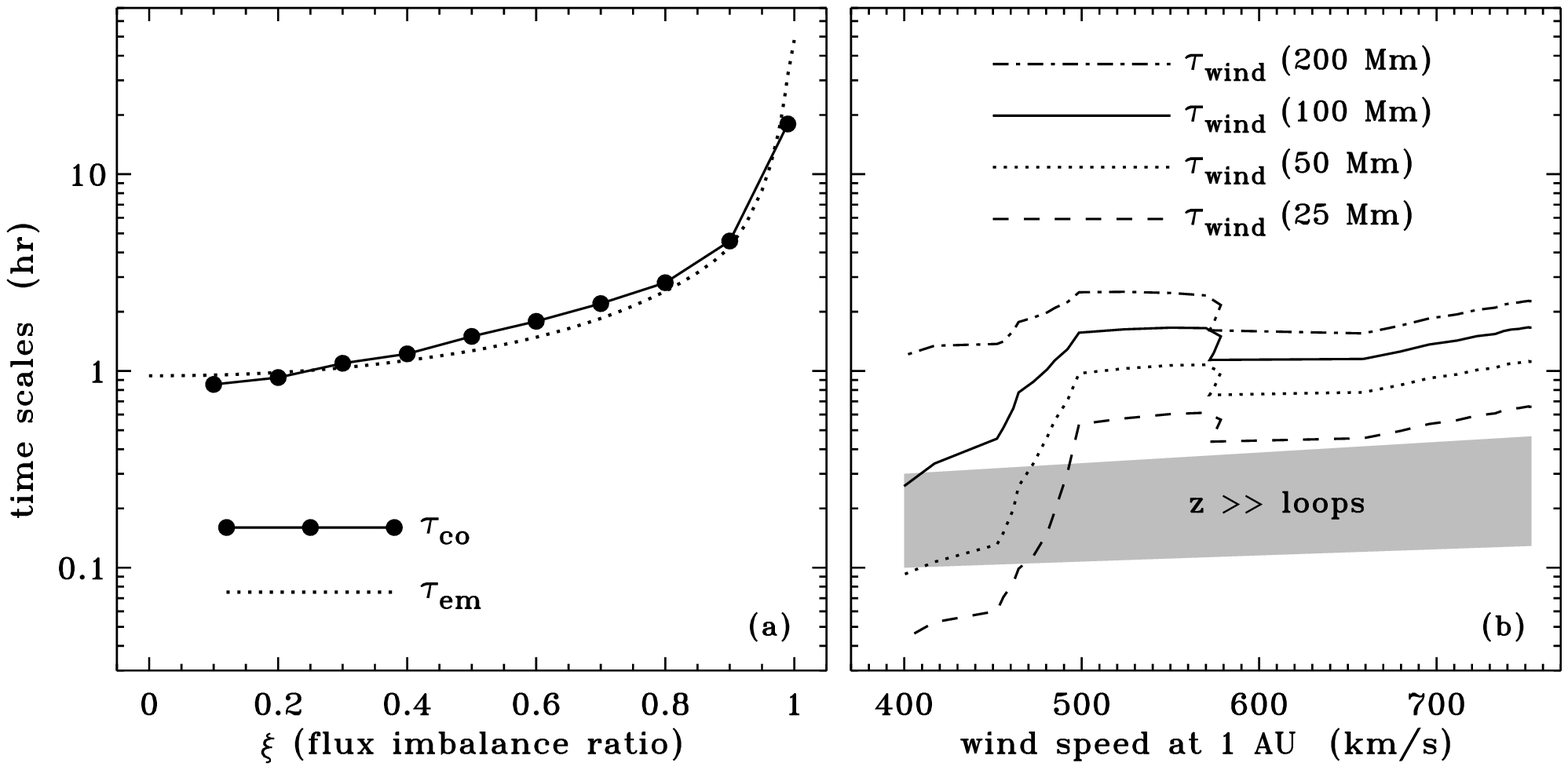}
\caption{Comparison of time scales for various models.
(a) For the Monte Carlo models of the magnetic carpet, the recycling
time for flux emergence (dotted curve) is compared with the time
scale for flux opening (filled circles and solid curve).
(b) For the \citet{CvB07} solar wind models, we plot acceleration
times $\tau_{\rm wind}$ up to heights of 25 Mm (dashed curve),
50 Mm (dotted curve), 100 Mm (solid curve), and 200 Mm
(dot-dashed curve) versus the outflow speeds at 1 AU.
Also shown is an approximate region of parameter space that corresponds
to upper heights $z$ that exceed 2--3 times the maximum heights of
closed loops in the corresponding BONES models (gray box).}
\label{fig10}
\end{figure*}

We studied the time evolution of magnetic topology in the BONES
simulations by following the opening and closing of flux tubes
from the beginning to the end of each time step.
For comparison, we also computed the recycling time scale for
flux to emerge from below the photospheric surface (see also
Section \ref{sec:3.2}).  We defined this quantity as
\begin{equation}
  \tau_{\rm em} \, = \, \frac{\langle B_{\rm abs} \rangle}{E}
  \,\, .
\end{equation}
For our models we took $\langle B_{\rm abs} \rangle$ from
Figure \ref{fig03} and $E$ from Equation (\ref{eq:Efit}), and we
found that the emergence time scale $\tau_{\rm em}$ tends to have
values around 1--2 hr \citep[see][]{Hg08}.
Regions of extreme flux imbalance undergo slower emergence, with
$\tau_{\rm em}$ exceeding 10--20 hr when $\xi \gtrsim 0.9$.
Figure \ref{fig10}(a) shows the $\xi$ dependence of this time scale.

Next we used the flux tubes traced in our simulations to investigate
the time scales for magnetic field evolution in the corona.
\citet{Cl05} performed a similar study in the limit of a balanced
field, with $\xi = 0$.
They computed a so-called {\em coronal flux recycling time} that is
meant to characterize a local rate of change of the coronal field.
This rate is driven both by reconnection and by topological
evolution of the complex ``hierarchical tree'' of footpoint domains
in the magnetic carpet.
Because changes in the coronal field can take place even without
any flux emergence or cancellation, \citet{Cl05} found that coronal
flux recycling times can be significantly shorter than
photospheric flux recycling times.
Changes in topological connections can occur purely as a result of
the horizontal motions of flux elements \citep[e.g.,][]{Ed09,Ed10a}.
\citet{Cl05} used an older photospheric flux recycling time of
$\tau_{\rm em} \approx 15$ hr, but they found that the coronal
flux recycling time can be as short as 1.4 hr.
When emergence and cancellation were suppressed, the coronal
time scale was approximately a factor of two larger ($\sim$3 hr)
but still much more rapid than $\tau_{\rm em}$.
Our models differ from those of \citet{Cl05} in that our photospheric
emergence time scale is now of the same order of magnitude as their
coronal recycling time scale.

Below we describe how we estimate how long it takes for {\em just
the open flux} to recycle itself in the corona.
We do not track the (possibly more numerous) changes in topology
that do not involve open flux tubes.
As summarized in Section \ref{sec:5.1}, over the course of a time
step some of the flux in the model is unaccounted for because it
has either emerged since the last time step or it has evolved
beyond recognition as the same flux element.
The remaining fraction of total absolute flux---i.e., that which
survives the time step unaltered---is called $\psi$, and
Figure \ref{fig09} shows how its mean value increases steadily
from about 0.82 to 0.95 as $\xi$ increases from 0 to 1.
A larger choice for the relative tolerance parameter $\delta$
would give a larger survival fraction $\psi$ (see below), but it
can be argued that too much tolerance would give rise to errors
in how flux tubes are identified and tracked.

For flux tubes that survive a time step relatively unchanged, we
compared the endpoints of the field lines traced at the beginning
and end of the time step.
The fluxes in these field lines are summed into four separate bins
that are defined by their connectivity.
The four bins correspond to four fractions of the total surviving
absolute flux:
$f_{\rm oo}$ (starts open, ends open),
$f_{\rm oc}$ (starts open, ends closed),
$f_{\rm co}$ (starts closed, ends open), and
$f_{\rm cc}$ (starts closed, ends closed).
Because the overall magnetic configuration of the system does not
vary strongly over a single time step, we found that
$f_{\rm oo} \approx f_{\rm open}$.
Also, the two fractions that denote change ($f_{\rm oc}$ and
$f_{\rm co}$) both tend to be small contributors to the total.
The mean values of $f_{\rm co}$ in the models tend to vary
between about 0.005 and 0.025, with the largest values occurring
for intermediate imbalance ratios of $\xi \approx 0.5$ and the
smallest values occurring at the extremes of $\xi = 0$ and 0.99.
We also note that the time averages of $f_{\rm co}$ and $f_{\rm oc}$
are always roughly equal to one another (as should be required for
a time-steady dynamical equilibrium).
For all 11 models, the time averages of these two fractions never
differ from one another by more than about 2\%.

At any one time, we define the amount of open (absolute) flux
density as $B_{\rm open} = f_{\rm open} B_{\rm abs}$.
We computed the instantaneous rate of opening in each time
step $\Delta t$ as
\begin{equation}
  \left( \frac{dB}{dt} \right)_{\rm co} \, = \,
  \frac{f_{\rm co} B_{\rm abs}}{\Delta t}  \,\, .
  \label{eq:dBdtco}
\end{equation}
Note that the above definition makes the implicit assumption that
$f_{\rm co}$ is the fraction of the {\em total} absolute flux
density in the simulation that opens up in one time step.
However, this fraction is only approximately $\psi$ times the
total absolute flux that opens up.
We assumed that the small fraction $(1 - \psi)$ that was not counted
contributes in the same way as the larger fraction $\psi$ that
was counted.  (This assumption is tested below.)
Thus, the mean time scale for the opening up of closed flux tubes is
\begin{equation}
  \tau_{\rm co} \, = \, \frac{\langle B_{\rm open} \rangle}
  {\langle (dB/dt)_{\rm co} \rangle} \, = \,
  \frac{\langle f_{\rm open} \rangle \, \Delta t}
  {\langle f_{\rm co} \rangle}  \,\, .
  \label{eq:tauco}
\end{equation}

Because the quantities $f_{\rm co}$ and $(dB/dt)_{\rm co}$ can be
quite variable from time step to time step, we realized that care
should be taken in computing the averages in Equation (\ref{eq:tauco}).
We ended up computing these averages in two independent ways.
First, we took simple arithmetic averages of the time series for
$(dB/dt)_{\rm co}$ and the other quantities.
Second, we integrated the rate defined in Equation (\ref{eq:dBdtco})
as a function of time to build up the {\em cumulative} amount of
flux density that is opened up over the course of the simulation.
This is a monotonically increasing function, but its increase with
time is intermittent because different amounts of flux are opened up
in each time step.
We fit the cumulative growth of opened flux density with a linear
function, and then used the slope of this linear fit as the mean
value of $(dB/dt)_{\rm co}$.
These two methods gave results that agreed with one another to
within about 10\%, and we used the latter technique for all values
reported below.

Figure \ref{fig10}(a) compares the above time scales with one another.
It is clear that $\tau_{\rm co} \approx \tau_{\rm em}$ in these models.
In other words, the time scale for the replacement of the photospheric
flux---via emergence from below---is the same as the time scale for
replacement of the open flux that feeds the solar wind.
At first glance, this appears to be a simple requirement for a
time-steady equilibrium, in the same way that
$f_{\rm co} \approx f_{\rm oc}$ is required to maintain a steady
state.
However, one can imagine situations where the rate of flux evolution
in the corona is not so strongly coupled to the emergence rate of
new flux from below \citep[e.g.,][]{Cl05}.
In our case, it is the use of potential fields---which are remapped
during each time step with no allowance for the storage of
free energy in the corona---that demands
$\tau_{\rm co} \approx \tau_{\rm em}$.
In other words, the BONES models reproduce the case of highly
efficient magnetic reconnection, where the corona ``processes''
the flux as quickly as it is driven (stressed or injected) from
below.
One can imagine that in a full MHD simulation the efficiency of
magnetic reconnection may not be so high, and thus the resulting
non-potential, current-filled corona should exhibit
$\tau_{\rm co} > \tau_{\rm em}$.

Note that Figure \ref{fig10}(a) does not show the value of
$\tau_{\rm co}$ for the $\xi = 0$ model.
As Equation (\ref{eq:tauco}) makes clear, in this case both the
numerator and denominator are numbers that should approach zero.
Ideally, there should be no open fields at all in a {\em perfectly}
balanced potential field.
The BONES models do in fact give slightly nonzero values for
$\langle f_{\rm open} \rangle$ and $\langle f_{\rm co} \rangle$,
but these are believed to be numerical artifacts arising from the
discrete nature of the field-line tracing technique.
We reiterate that we do not compute the time scale for {\em all} of
the coronal flux to be recycled.
That recycling time should be nonzero even for the balanced
$\xi = 0$ model \citep{Cl05}.
In all models with $\xi \ll 0$, the full coronal recycling time
is likely to be significantly shorter than $\tau_{\rm co}$.

In order to study the dependence of our results on the assumptions
made about flux-tube identification, we varied the threshold
flux identification parameter $\delta$ away from its default value
of 0.1, in a range between 0 and 0.5.
This parameter sets the relative tolerance for the classification
of evolving flux elements over a time step.
Table 1 shows several resulting parameters of the test simulations,
which were all performed for $\xi = 0.4$.
As we expected, the flux survival fraction $\psi$ increases
monotonically with increasing $\delta$.
However, there does not seem to be any definitive trend with
$\delta$ in the fraction of flux that opens up ($f_{\rm co}$),
the related time scale for flux opening ($\tau_{\rm co}$), or
the energy flux released by reconnection into open-field regions
($\langle F_{\rm co} \rangle$, see Section \ref{sec:5.5}).
This suggests that the topological changes resulting from flux-tube
opening are adequately resolved in the simulations.
Thus, we retain the standard value $\delta = 0.1$ for the remainder
of the paper.

\begin{deluxetable}{lcccc}
\label{tab01}
\tablecaption{Variation of Mean Magnetic Properties
($\xi = 0.4$ model) with $\delta$}
\tablewidth{0pt}
\tablehead{
\colhead{$\delta$} &
\colhead{$\psi$} &
\colhead{$\langle f_{\rm co} \rangle$} &
\colhead{$\tau_{\rm co}$} &
\colhead{$\langle F_{\rm co} \rangle$} \\
\colhead{} &
\colhead{} &
\colhead{} &
\colhead{(hr)} &
\colhead{(erg cm$^{-2}$ s$^{-1}$)}
}

\startdata

0.00 & 0.759 & 0.0199 & 1.403 & $1.50 \times 10^{4}$ \\
0.10\tablenotemark{a} & 0.839 & 0.0220 & 1.222 & $1.77 \times 10^{4}$ \\
0.25 & 0.901 & 0.0219 & 1.273 & $1.65 \times 10^{4}$ \\
0.38 & 0.926 & 0.0245 & 1.160 & $1.80 \times 10^{4}$ \\
0.50 & 0.938 & 0.0226 & 1.233 & $1.70 \times 10^{4}$ \\

\enddata
\tablenotetext{a}{Standard value used in all other models
discussed below.}

\end{deluxetable}

It is worthwhile to compare the time scale for flux opening to the
time scale for solar wind acceleration along the open flux tubes.
If a significant amount of solar wind plasma flows out during the
time it takes the open field to reorganize itself via reconnection,
then the reconnection processes themselves probably {\em are not
responsible} for producing the majority of the solar wind.
The RLO idea depends on the plasma in open flux tubes coming from
the opening up of closed loops.
Thus, we want to determine whether or not a large amount of mass
accelerates out in the open flux tubes over the time it would take
for significant mass to be processed via loop-opening.

The time scale for wind acceleration from a lower height
$z_{\rm TR}$ in the solar transition region (TR) to an arbitrary
upper height $z$ is
\begin{equation}
  \tau_{\rm wind} (z) \, = \, \int_{z_{\rm TR}}^{z}
  \frac{dz'}{u(z')}  \,\, ,
  \label{eq:tauwind}
\end{equation}
where $u(z)$ is the radial wind speed.
The TR was chosen as the height to start the integration because
that is where the mass flux of the wind is thought to be determined
\citep[see, e.g.,][]{Ha82,Wi88,HL95}.
We used the one-fluid solar wind models of \citet{CvB07} to compute
$\tau_{\rm wind}$, and we defined $z_{\rm TR}$ as the height at
which the temperature in a wind model first reaches $10^{5}$ K.

Figure \ref{fig10}(b) shows the wind acceleration time scales for
several representative upper heights $z$, and for a range of models
of the fast and slow wind that have speeds at 1 AU between 400
and 750 km s$^{-1}$ \citep[see][]{CvB07}.
Two side-by-side plots are necessary in Figure \ref{fig10}
because there is not a unique one-to-one correspondence between
the flux imbalance ratio $\xi$ and the wind speed at 1 AU.
We do know, however, that there is some association between
slow wind streams and QS regions on the surface ($\xi \approx 0$)
and between fast wind streams and CH regions on the surface
($\xi \approx 1$).
Thus, the overall left-to-right variations in the two panels can
be roughly associated with one another.

The slow solar wind models shown in Figure \ref{fig10}(b) have
the shortest acceleration time scales.
Given Equation (\ref{eq:tauwind}), this is potentially counterintuitive.
However, we note that the slow wind models from \citet{CvB07}
often have local maxima in $u(z)$ of order 100 km s$^{-1}$ in
the low corona that are not present in the more steadily
accelerating fast wind models \citep[see also Figure 7a of][]{Cr10}.
These regions correspond to enhanced magnetic fields that were
included to simulate open fields at the edges of streamers and
active regions.
Observations are beginning to show hints of such rapid outflows
as well \citep{Ha08,Su10,Br10}.

When comparing the time scales for flux opening and solar wind
acceleration, we can use the loop heights illustrated in
Figure \ref{fig07} as an order-of-magnitude guide for the maximum
height $z$ to use when computing $\tau_{\rm wind}(z)$.
For example, when parcels of solar wind exceed a height that is
2--3 times $H_{95}$, it can be safely assumed that the wind has
left behind virtually all interactions with closed loops
and should be considered to be freely accelerating.
This allows us to compare the time scales between panels in
Figure \ref{fig10} for the two general types of solar wind:
\begin{enumerate}
\item
For slow wind streams rooted in balanced QS regions (i.e.,
$\xi \approx 0$), the height at which the wind flows
``free and clear'' of loops is of order 50--100 Mm.
Figure \ref{fig10}(b) shows that this height corresponds to
$\tau_{\rm wind} \approx 0.1$--0.3 hr.
This is a shorter time scale than the representative flux-opening
time $\tau_{\rm co} \approx 1$ hr that corresponds to the left side
of Figure \ref{fig10}(a), but it is still of the same order of
magnitude.
Thus, it is possible that RLO processes could be important for
slow wind acceleration.
\item
For fast wind streams rooted in unbalanced CH regions (i.e.,
$\xi \approx 1$), the height corresponding to 2--3 times $H_{95}$
is only of order 5--15 Mm.
The fast wind accelerates to this range of heights in less than
about 0.3 hr, but the flux-opening recycling time in coronal holes
can be as long as 3--10 hr.
This is a larger discrepancy than in the case of the slow wind,
and it implies that it is unlikely that RLO processes are important
in accelerating the bulk of the fast wind.
(Of course, it still may be the case that RLO processes produce
a highly intermittent or episodic injection of mass and energy into
the fast wind in coronal holes---just not enough to affect the
majority of the accelerating plasma.
The polar jets discussed further in Section \ref{sec:6} may be a
prime example of this intermittency.)
\end{enumerate}
The gray box in Figure \ref{fig10}(b) shows the approximate range of
wind acceleration time scales that correspond to maximum heights
$z$ exceeding about 2--3 times $H_{95}$ as discussed above.
The shape of the gray region is roughly independent of wind speed
and $\xi$.
This is because, as one goes from left to right in the plot, the
increase in $\tau_{\rm wind}$ (for constant $z$) is offset by
the fact that the relevant value of $z$ decreases (because 
$H_{95}$ decreases; see Figure \ref{fig07}).

Finally, we reiterate that the values of $\tau_{\rm co}$ shown in
Figure \ref{fig10}(a) are likely to just be lower limits to the
actual time scales of flux-opening.
As discussed above, our models assume a succession of potential
fields that are consistent with the assumption of rapid magnetic
reconnection.
If the true MHD state of the corona exhibits less efficient
magnetic reconnection, then the photospheric footpoint stressing
will build up non-potential fields and current sheets in the
corona and thus give rise to larger net values of $\tau_{\rm co}$.
In this case, it is even more certain that
$\tau_{\rm co} \gg \tau_{\rm wind}$, and our conclusion that
RLO processes are unimportant in accelerating the solar wind
is strengthened.

\subsection{Poynting Flux in Emerging Bipoles}
\label{sec:5.4}

Our primary reason for constructing the BONES simulations was to
estimate how much energy is deposited into the solar wind by the
evolving magnetic carpet.
First, though, it is necessary to compute how much magnetic energy
is being injected into the system from below the photosphere.
It is not obvious that all (or even most) of this energy is able to
be converted into forms that supply heat or momentum to the
accelerating solar wind.
Since, on small scales, much of the injected magnetic energy is in
the form of compact bipoles, it may be difficult for much of this
energy to become ``liberated'' into the open-field regions
when these bipoles evolve and interact with one another.
Thus, in this section we discuss the total magnetic energy that is
potentially available, and in the following section we estimate
what fraction of it is actually released by reconnection into the
open-field regions.

The relevant quantity to compute when considering the rate of
injection of magnetic energy from below the photosphere is the
Poynting flux, which is defined as
\begin{equation}
  {\bf S} \, = \, \frac{c}{4\pi} {\bf E} \times {\bf B}
  \, \approx \, -\frac{1}{4\pi} \left[ \left(
  {\bf v} \times {\bf B} \right) \times {\bf B} \right]
  \,\, ,
\end{equation}
and where the latter approximation assumes the ideal condition of
MHD flux freezing.
In the Cartesian system studied in this paper, the most relevant
component of the Poynting flux is the $z$ component, with
\begin{equation}
  S_{z} \, = \, \frac{1}{4\pi} \left[ B_{\perp}^{2} v_{z} -
  ({\bf v}_{\perp} \cdot {\bf B}_{\perp}) B_{z} \right]
  \label{eq:Sz}
\end{equation}
where ${\bf B}_{\perp}$ and ${\bf v}_{\perp}$ are the components
of the magnetic field and velocity in the horizontal ($x$--$y$)
plane.
The two terms on the right-hand side of Equation (\ref{eq:Sz})
represent components associated with flux emergence and surface
flows, respectively.
For simplicity, though, in the remainder of this section we will
endeavor only to estimate the overall magnitude $S$ of the Poynting flux.
This gives a reliable upper limit that is independent of the
adopted geometry and topology of the emerging flux elements.

Observationally, the Poynting flux can be estimated from various
measured proxies \citep[e.g.,][]{We09}, but there exist ambiguities
in the data that give rise to significant uncertainties.
\citet{Fi99} estimated the magnitude of ${\bf S}$ to be about
$5 \times 10^{5}$ erg cm$^{-2}$ s$^{-1}$ in source regions of
the solar wind.
\citet{MG10} used vector magnetic fields measured by
{\em{Hinode}}/SOT to estimate that small-scale emerging loops
provide something like $10^6$ to $2 \times 10^{7}$
erg cm$^{-2}$ s$^{-1}$ to the low chromosphere in quiet regions.

\begin{figure*}
\epsscale{0.93}
\plotone{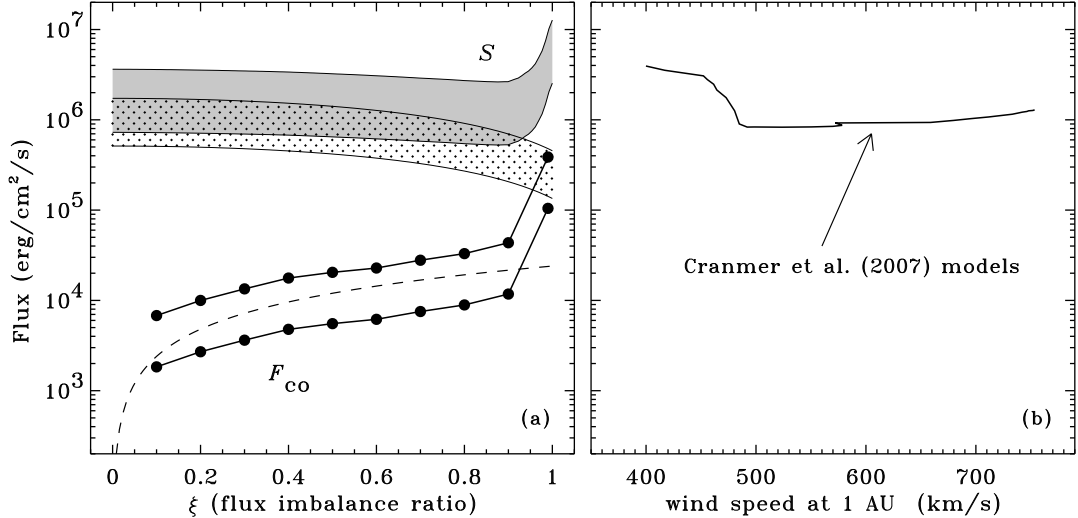}
\caption{Comparison of energy fluxes for various models.
(a) Estimated flux $\langle F_{\rm co} \rangle$ in loop-opening
events (filled circles and solid curves) computed with two choices
for $\theta_{\rm L} C_{\rm L}$.
Also shown are approximate Poynting fluxes $S$ for photospheric flux
emergence, with the dotted region showing estimates from Equation
(\ref{eq:Salt1}) and the gray region showing estimates from
Equation (\ref{eq:Salt2}).
The dashed curve shows a linear scaling
$\langle F_{\rm co} \rangle \propto \xi$.
(b) Total dissipated solar wind energy flux $F_{\rm wind}$ from
the WTD models of \citet{CvB07}.}
\label{fig11}
\end{figure*}

We estimated the magnitude of the Poynting flux for the
Monte Carlo models developed above in two independent ways.
Figure \ref{fig11}(a) shows that the two methods gave rise to
similar ranges of Poynting flux (both of order $10^{6}$
erg cm$^{-2}$ s$^{-1}$) with a relatively weak dependence on $\xi$.
These two methods are described below.

First, we note that the emergence rate $E$ (Equation (\ref{eq:Efit}))
already describes how much magnetic flux is driven up from below the
photosphere, per unit area and per unit time (i.e., its units are
Mx cm$^{-2}$ s$^{-1}$).
What we want to know is how much magnetic energy emerges, in units
of erg cm$^{-2}$ s$^{-1}$.
Thus, if we can relate the flux in an emerging bipole to its
magnetic energy, we can convert easily from $E$ to $S$.
Treating a pair of equal-and-opposite emerging flux elements as an
an ideal (but partially submerged) magnetic {\em dipole,} we can
specify its field strength as
\begin{equation}
  B \, = \, \frac{\Phi_{i} D}{2\pi r^{3}} \sqrt{1 + 3 \cos^{2} \theta}
  \,\, ,
\end{equation}
where $\Phi_i$ is the absolute flux in each pole,
$D$ is the horizontal separation between the two poles, $r$ is
the distance measured from the center of the dipole, and $\theta$
is the polar angle measured from the (horizontal) dipole axis.
Assuming the dipole is submerged at a depth $|z_{i}|$, it is possible
to integrate the magnetic energy $U_{\rm mag}$ over the full coronal
volume $V$ (i.e., over all $x$ and $y$, and all $z > 0$) analytically.
We thus found
\begin{equation}
  U_{\rm mag} \, = \, \int dV \, \frac{B^2}{8\pi} \, = \,
  \frac{\Phi_{i}^{2} D^{2}}{128 \pi^{2} |z_{i}|^3}  \,\, .
\end{equation}
Note that the magnetic energy above the photosphere is extremely
sensitive to the submerged depth $|z_{i}|$.
Once the magnetic energy due to a given bipole is known, we can
estimate the magnitude of the Poynting flux as
\begin{equation}
  S \, \approx \, E \, \frac{\langle U_{\rm mag} \rangle}
  {\langle \Phi \rangle}
  \label{eq:Salt1}
\end{equation}
where the angle brackets denote the properties of the ``average''
emerging bipole as discussed in Section \ref{sec:3.2}.
Figure \ref{fig11}(a) shows this quantity for the 11 models
as a function of $\xi$,
and for two reasonable choices of $|z_{i}|$ (0.8 and 1.2 Mm).
For typical values of $E = 10^{-3}$ Mx cm$^{-2}$ s$^{-1}$,
$\langle \Phi \rangle = 9 \times 10^{18}$ Mx,
$D = 6.8$ Mm, and $|z_{i}| = 1$ Mm, we find that
$S \approx 8 \times 10^{5}$ erg cm$^{-2}$ s$^{-1}$.

The second way to estimate $S$ was proposed by \citet{Fi99}.
Here, we compute the total magnetic energy in the system
(per unit surface area) and divide it by the flux recycling time.
In other words,
\begin{equation}
  S \, \approx \, \frac{1}{\tau_{\rm em}} \int dz \, \frac{B^2}{8\pi}
  \,\, .
  \label{eq:Sdef2}
\end{equation}
Here, the value of $B$ at the photospheric surface is essentially
the time-averaged absolute flux density (i.e.,
$B_{\odot} \approx \langle B_{\rm abs} \rangle$).
It is the height-dependence of $B$, for $z>0$, that is the major
source of uncertainty in evaluating Equation (\ref{eq:Sdef2}).
However, it is straightforward to follow \citet{Fi99} and assume
a vertical falloff that depends on a power of heliocentric radius.
Thus,
\begin{equation}
  B \, \approx \, B_{\odot} \left( \frac{R_{\odot}}{r}
  \right)^{n}
\end{equation}
(where $r = z + R_{\odot}$), and then
\begin{equation}
  S \, \approx \, \frac{B_{\odot}^{2} R_{\odot}}
  {8\pi \tau_{\rm em} (2n-1)}  \,\, .
  \label{eq:Salt2}
\end{equation}
At large distances above the photosphere, the exponent $n$
approaches a value of 2, but it is believed to take on larger
values closer to the surface \citep[see, e.g.,][]{Ba98}.
For a typical value of $B_{\odot} = 4$ G and
$\tau_{\rm em} = 1$ hr, we can estimate an upper limit on $S$
by assuming $n=2$, and thus obtain $S = 4 \times 10^{6}$
erg cm$^{-2}$ s$^{-1}$.
For a more realistic coronal value of $n \approx 8$, we have
$S \approx 6 \times 10^{5}$ erg cm$^{-2}$ s$^{-1}$.
Figure \ref{fig11}(a) shows how $S$ varies as a function of $\xi$
when the modeled variations in $\langle B_{\rm abs} \rangle$ and
$\tau_{\rm em}$ are used, and when the two above values of $n = 2$
and 8 are assumed to define the lower and upper limiting cases.
Given the uncertainties, the two alternate methods of estimating
$S$ give numerical values that are quite consistent with one another.

\subsection{Energy Release in Loop-Opening Events}
\label{sec:5.5}

We used the output of the BONES simulations to estimate the amount
of energy released by magnetic reconnection for cases of closed
flux tubes turning into open flux tubes (and vice versa).
It is important to note that there are also expected to be many
{\em other} sites of reconnection and energy release that do not
involve open flux tubes.
For example, in a balanced QS region there may be a large number of
small-scale ``footpoint-swapping'' events that start with a
configuration of closed loops and end with a slightly different
topological distribution of closed loops \citep{Pr02,Cl05}.
In this paper, we explicitly ignore the energy release in the
closed--closed events in order to focus on only the subset of
events that can input mass and energy into the solar wind.

The basic geometrical picture for a flux-opening event is the
``anemone'' type structure that is believed to exist at the
footpoints of many X-ray bright points, coronal jets, and polar
plumes \citep[e.g.,][]{Sy82,Sh92,Sh07,Fp09,ST09}.
In this picture, a small bipolar magnetic field either emerges or
advects into the presence of a larger-scale open field.
Magnetic reconnection is believed to occur roughly above the
end of the bipole with the opposite polarity as the open field
\citep{Ed09,Ed10a}.
The newly opened flux may take the form of a jet or plume
\citep{Wa98}, and the newly closed flux may ``subduct'' and
provide heating to the underlying chromosphere \citep{Gg08}.
In one of these interchange-reconnection type events, the amount
of closed magnetic flux that opens up should be the same as the
amount of pre-existing open flux that becomes closed
(i.e., $f_{\rm co} \approx f_{\rm oc}$).

Because we model the evolution of the coronal magnetic field as a
succession of potential fields (see Section \ref{sec:2}), we use
the quasi-static ``minimum current corona'' (MCC) model to estimate
the energy loss due to reconnection \citep{Lg96,LK99,BL06}.
In this model, the mean energy flux released in closed-to-open
reconnection events is proportional to the rate $(dB/dt)_{\rm co}$
at which magnetic flux is opened up (see Equation (\ref{eq:dBdtco})).
For our simulations, we derived the MCC energy flux to be
\begin{equation}
  F_{\rm co} \, = \,
  \theta_{\rm L} C_{\rm L} \frac{\Phi_1}{\langle d \rangle}
  \left| \frac{dB}{dt} \right|_{\rm co} \,\, ,
  \label{eq:Fco}
\end{equation}
where $\Phi_{1}$ is the mean absolute flux per element,
$\langle d \rangle$ is the mean separation between elements in
the simulation, and $\theta_{\rm L}$ and $C_{\rm L}$ are
dimensionless constants.
The Appendix presents a detailed derivation of
Equation (\ref{eq:Fco}) for anemone-type reconnection events,
including a discussion of the most likely numerical values for
$\theta_{\rm L}$ and $C_{\rm L}$.

It is important to clarify that the energy flux given by
Equation (\ref{eq:Fco}) is meant to be an order-of-magnitude
representation of the magnetic ``free energy'' released by
reconnection.
The MCC model depends on an estimate of the current that builds up
and is dissipated along an idealized separator, and truly
non-potential MHD simulations are needed to verify whether these
estimates are valid.
Also, the MCC model does not specify how the energy is
partitioned into other forms such as thermal energy, bulk
kinetic energy, waves, MHD turbulence, and energetic particles.
Determining this partitioning is a complex problem---one definitely
beyond the scope of this paper---that often requires the use of
fully kinetic simulations.
However, it has been found that many forms of particle energization
that occur rapidly and locally in reconnection regions may
eventually become unstable to dissipation that randomizes the
velocity distributions \citep{Bh04,FM06,Ya07}.
Thus, much of the energy that initially goes into, e.g., waves or
supra-Alfv\'{e}nic beams may end up released in the form of heat.
This will be our implicit assumption when comparing
$F_{\rm co}$ with the energy fluxes required to heat the corona
and accelerate the solar wind along open flux tubes.

Figure \ref{fig11}(a) shows the time-averaged quantities
$\langle F_{\rm co} \rangle$ for 10 of the standard BONES models
as a function of $\xi$ (excluding the case $\xi=0$).
See below for a discussion of how $F_{\rm co}$ varies in time.
The lower and upper sets of points were computed by assuming the
product of the two dimensionless constants
$\theta_{\rm L} C_{\rm L}$ to be 0.003 and 0.011, respectively
(see the Appendix).
For nearly all of the models, $\langle F_{\rm co} \rangle$ is
significantly smaller than the available Poynting flux $S$.
For the lowest values of $\xi$, the resulting ``efficiency''
of energy release in open-field regions (i.e.,
$\langle F_{\rm co} \rangle / S$) may be as low as 0.001--0.01.
This means that in QS regions, only a tiny fraction of the
magnetic energy that enters the system ends up being available
for driving the solar wind via RLO processes.

For most values of $\xi$, the computed values of
$\langle F_{\rm co} \rangle$ are significantly lower than the
canonical heat fluxes (i.e., $3 \times 10^{5}$ to $10^6$
erg cm$^{-2}$ s$^{-1}$) that \citet{WN77} estimated are needed
to maintain QS and CH regions on the Sun.
However, for the most unbalanced CH regions ($\xi \gtrsim 0.95$)
the modeled energy fluxes do appear to approach both the empirically
required heating rates and the empirically constrained Poynting
fluxes.
Observed coronal holes, however, exhibit values of $\xi$ over a
much wider range of values \citep{WS04,Ab09}, so the models still
have a problem with explaining CH coronal heating in general.

Figure \ref{fig11}(a) also shows a curve that represents a linear
dependence with the flux imbalance ratio; i.e.,
$\langle F_{\rm co} \rangle \propto \xi$.
For $0.2 \leq \xi \leq 0.9$, this linear relationship
appears to fit the variation in the modeled energy fluxes.
Because we also know that $f_{\rm open} \propto \xi$ (see
Figure \ref{fig09}), this tells us that the heating rate in
flux-opening events is roughly proportional to how much of the
time-averaged magnetic field remains open.

As was done in Section \ref{sec:5.3} above, we can also compare
the results from the BONES simulations with earlier models of solar
wind acceleration along open flux tubes.
We would like to assess how much energy flux needs to be deposited
in open-field regions in order to produce the solar wind.
We used the one-fluid WTD-type models of \citet{CvB07} to estimate
this quantity.
These models involved finding a self-consistent description of the
volumetric heating rate $Q = |\nabla \cdot {\bf F}|$ (in units of
erg cm$^{-3}$ s$^{-1}$) that was able to maintain time-steady
corona and solar wind.
In order to derive the total energy flux $|{\bf F}|$ that was
dissipated in one of these models, we had to integrate over the
entire radial grid, which extended from the photosphere to the
heliosphere.
The \citet{CvB07} models were computed along magnetic flux tubes
that have a radially varying cross-sectional area $A_{\rm tube}(z)$.
Thus, the radial integral of the product $Q A_{\rm tube}$ gives
the total power dissipated (in erg s$^{-1}$) in a flux tube.
To express this quantity as an energy flux and compare it to the
quantities shown in Figure \ref{fig11}(a), we normalized the
area function $A_{\rm tube}(z)$ to the area of the simulation box
($A = [200 \, \mbox{Mm}]^{2}$) at a height corresponding to the
low corona, at which the supergranular funnels have expanded to
fill the ``canopy'' volume.
For the \citet{CvB07} models, this height corresponds to
$z \approx 0.04 \, R_{\odot} \approx 28$ Mm.
Then the energy flux can be computed by dividing the total power
by the box area $A$, and
\begin{equation}
  F_{\rm wind} \, = \,
  |{\bf F}| \, = \, \frac{1}{A} \int_{0}^{\infty} dz \,\,
  A_{\rm tube}(z) \, Q(z)  \,\, .
\end{equation}
Figure \ref{fig11}(b) shows how $F_{\rm wind}$ depends on
the wind speed at 1 AU for the same models that were shown
in Figure \ref{fig10}(b).
We point out that \citet{Fi99} was correct to conclude that the
energy flux {\em needed} to accelerate the solar wind is of
the same order of magnitude as the emerging Poynting flux $S$
\citep[see also][]{Le82,SM03}.
However, Figure \ref{fig11}(a) shows that RLO-type flux-opening
events do {\em not} appear to be able to release the required energy
flux into the open flux tubes.

\begin{figure}
\epsscale{1.09}
\plotone{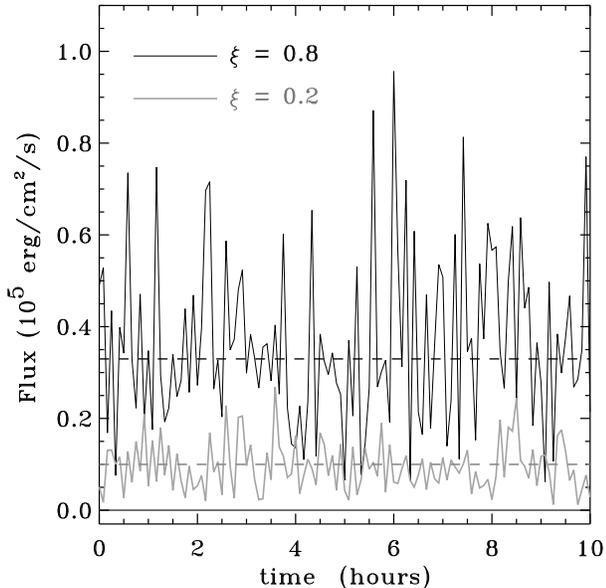}
\caption{Time evolution of the energy flux $F_{\rm co}$ released
by reconnection into open-field regions, for BONES models having
$\xi = 0.2$ (gray curves) and $\xi = 0.8$ (black curves).
Time averages for both cases are denoted by dashed lines.}
\label{fig12}
\end{figure}

A key result of many coronal heating models---including the MCC
models of \citet{Lg96}---is that the energy dissipation process
should be highly intermittent.
This occurs in the BONES simulations as well.
Figure \ref{fig12} shows a snapshot of the time dependence of
the quantity $F_{\rm co}$ for the $\xi = 0.2$ and 0.8 models.
These heating rates were computed with the upper-limit value of
the product $\theta_{\rm L} C_{\rm L} = 0.011$.
Thus, the time averages of these quantities correspond to the
upper set of solid points in Figure \ref{fig11}(a).
For the majority of the models ($0.2 \leq \xi \leq 0.9$) the
standard deviation of $F_{\rm co}$ is approximately half of its
mean value.
For the extreme models with the lowest and highest values of $\xi$,
the standard deviations increase to be about equal to their means.
Such a scaling would be expected if the energy fluxes were sampled
from an {\em exponential distribution} similar in form to that of
the emerging bipole fluxes (Equation (\ref{eq:probE})).
In any case, the variability of the predicted heating rates may
be just as useful as the mean values when attempting to
distinguish between different coronal heating models
\citep[see, e.g.,][]{Pk88,WG00,BV07}.

It is worthwhile to list some of the ways in which the above
models may be incomplete or incorrect.
For example,
\begin{enumerate}
\item
The assumption of a succession of potential fields is likely
to limit the verisimilitude of the models.
It is clear that time dependent, three-dimensional MHD
models---which contain currents, resistivity, and finite-pressure
effects---would shed more light on the dynamics and energetics of
this system.
If the gas pressure in localized reconnection regions begins to
exceed the magnetic pressure (i.e., $\beta \gtrsim 1$),
there may be additional ways for the flux tubes to ``break open''
that were not accounted for here.
\item
Even within the confines of a succession of potential fields,
the assumptions of the MCC model may be too simplistic.
For example, it is known that in three-dimensional reconnection there
are both spatial and temporal variations of the current {\em along}
separators, which our implementation of MCC does not include
\citep[e.g.,][]{GP05,Pn10}.
\item
Our assumption of $\theta_{\rm L} = 1$ in Equation (\ref{eq:Fco})
may be too large, and thus our resulting estimate for the energy
flux released by reconnection may be too high.
\item
The simple three-pole magnetic geometry discussed in the Appendix
did not consider realistic asymmetries in either the footpoint
locations or the magnitudes of the flux sources.
When such asymmetries are included \citep{AP10}, the resulting
range of values for $C_{\rm L}$ would likely be different.
It is unclear whether $C_{\rm L}$ would be larger or smaller than
the values estimated in the Appendix.
\item
The use of the mean flux element separation $\langle d \rangle$
in Equation (\ref{eq:Fco}) is only a rough approximation.
Since there may be significant energy release when one flux element
gets very close to another, it may be better to use a mean distance
that is smaller than $\langle d \rangle$.
In that case, our estimate for the heating rate could be too low.
\item
As we mentioned in Section \ref{sec:5.1} above, many of the flux
tubes that are classified as ``open'' may in fact be closed in the
form of hydrostatic helmet streamers.
In reality, then, the energy flux that escapes out into the solar
wind could be even lower than the values of
$\langle F_{\rm co} \rangle$ that were shown in
Figure \ref{fig11}(a).
It is also possible that large-scale interchange reconnection could
eventually open up these flux tubes \citep{Wa00,Fi05,An10,Ed10a}, but
modeling those processes is beyond the scope of this paper.
\end{enumerate}
Roughly speaking, there appear to be just as many reasons why our
results for the rates of RLO heating and flux-opening may be
overestimates as there are reasons why they may be underestimates.
Despite the approximate nature of these models, however, we
believe that the main result (i.e., $\langle F_{\rm co} \rangle
\ll S$ for most values of $\xi$) is not likely to be wrong by
many orders of magnitude. 

\section{Discussion and Conclusions}
\label{sec:6}

The primary aim of this paper was to begin testing the
conjecture that the opening up of closed flux in the Sun's
magnetic carpet is responsible for driving the solar wind.
First, we created Monte Carlo simulations of the complex
photospheric sources of the solar magnetic field.
The resulting time-averaged properties of the models appeared to
agree well not only with observations of the flux density and the
flux imbalance ratio, but also with observed probability
distributions for the flux elements and autocorrelation functions
of the field strength.
A supergranular pattern of network magnetic concentrations
appeared spontaneously in the models, despite the lack of any
imposed supergranular motions.
Then, armed with some degree of confidence that the model photosphere
is an adequate reflection of reality, we then computed the coronal
magnetic field.
Assuming that the coronal field evolves as a succession of
potential-field extrapolations, we were able to estimate both the
time scales and energy fluxes associated with RLO-type
flux-opening events.

From the simulations, we found that the Poynting flux in emerging
magnetic elements (which could be a proxy for the maximum energy
flux available for coronal heating) is typically around
$10^{6}$ erg cm$^{-2}$ s$^{-1}$.
However, for quiet regions ($\xi \ll 1$), only a tiny fraction
of the available Poynting flux was found to be released in 
flux-opening events via magnetic reconnection.
A similar situation was found to exist in mixed-polarity regions
that can correspond to either quiet Sun or coronal holes
($\xi \lesssim 0.8$).
For the most unbalanced coronal hole regions ($\xi \approx 1$),
the fraction of Poynting flux released in flux-opening events may
approach unity.
In these regions, however, the time scale for flux opening was
found to be significantly longer than the solar wind travel time
from the coronal base to heights far above the tops of loops.
Thus, it appears that a significant amount of mass accelerates
out into the solar wind over the time that it would take for the
plasma to be processed via RLO type mechanisms.
From the above estimates of time scales and MCC energetics,
we conclude that {\em it is unlikely that
the solar wind is driven by reconnection and loop-opening
processes in the magnetic carpet.}

Despite the negative conclusion regarding the solar wind as a whole,
we believe that the physical processes modeled in this paper are
likely to be relevant in many other ways.
For example, it is possible that more can be learned about the
energetics of {\em polar jets} with the methodology developed here.
Soft X-ray observations can be used to estimate the energy flux
released due to jet eruptions.
These jets are believed to span several orders of magnitude in
the total amount of energy released; i.e., between about $10^{26}$
and $10^{29}$ erg \citep{Sh98,Ch08,Pt09,Mo10}.
Let us take a canonical value of
$E_{\rm jet} \approx 4 \times 10^{28}$ erg from the model of
\citet{Sh98}.
Recently, \citet{Sv07} identified 104 jets with the {\em Hinode}
X-Ray Telescope (XRT) over a time span of 44 hours in a polar coronal
hole, which gives a mean time between jets (for the observed area)
of $\tau_{\rm jet} \approx 1500$ s.
The area examined by \citet{Sv07} was approximately the ``front half''
of the polar cap, viewed from the side, which extended down
to about {25\arcdeg} colatitude and thus covered about
$A_{\rm jet} \approx 1.5 \times 10^{21}$ cm$^{2}$.
Thus, we estimate the mean energy flux released in jets to be
$F_{\rm jet} \approx E_{\rm jet} / (A_{\rm jet} \tau_{\rm jet})
\approx 2 \times 10^{4}$ erg cm$^{-2}$ s$^{-1}$.
This agrees reasonably well with the predicted energy fluxes
(for $\xi \approx 0.6$--0.9) shown in Figure \ref{fig11}(a).

The flux-opening events modeled in this paper may also be relevant
to understanding the small eruptions seen in quiet regions
\citep{In09,Sj10} that may be related to coronal mass ejections
(CMEs).
However, it is not guaranteed that every jet-like eruption
observed in the corona releases material that accelerates up
into the solar wind.
There is observational evidence that at least some coronal jets
contain plasma that falls back down because it failed to
reach the escape speed \citep{Bk08,Sc09}.
This may put some jets into the same category as spicules, which
are known to carry orders of magnitude more mass up (and down)
than is needed to feed the solar wind \citep[e.g.,][]{St00,DP09}.

A potentially valuable set of observational diagnostics of the
processes discussed in this paper are the elemental abundances and
ionization states of different particle species that escape into
the solar wind \citep{Zu07}.
The closed-to-open reconnection events that we have modeled may
inject some plasma with a distinctly ``closed'' composition
signature into regions that have signatures otherwise dominated by
flux tubes that remain open.
It is worth noting, however, that there remains disagreement about
exactly what kinds of abundance and ionization signatures signal
the presence of closed loops, and which do not.
\citet{CvB07} showed that a range of WTD-type open-flux-tube models
can produce values of the commonly measured O$^{7+}$/O$^{6+}$ and
Fe/O ratios that agree reasonably well with in~situ measurements
\citep[see also][]{Pu10}.
Thus, we question the popular assertion that the charge-state and
first-ionization-potential (FIP) properties measured in the slow
solar wind can only be explained by the injection of plasma from
closed-field regions on the Sun.

Whether or not the solar wind energy budget is accounted for by
RLO processes, the inherent {\em variability} in the magnetic carpet
is likely to cause some kind of MHD fluctuations to propagate up
into the corona.
The response of the coronal field to the evolving footpoints
may result in Alfv\'{e}n waves with periods of order
$\tau_{\rm em} \approx \tau_{\rm co}$ (see Figure \ref{fig10}).
In fact, \citet{Ho90,Ho08} suggested that ``flux cancellation''
events in the corona may be the most likely source of the long-period
(i.e., 0.5--10 hour) Alfv\'{e}n waves that dominate in~situ
measurements.
The statistical properties of these low-frequency fluctuations
may also be consistent with an origin in the motions of coronal
field-line footpoints \citep{MG86,GJ04,Nc09}.

In order to further test the applicability of RLO-type processes
to accelerating the solar wind, the models need to evolve beyond the
approximate potential-field ``skeleton'' and to incorporate MHD
effects.
Multi-dimensional MHD simulations \citep[e.g.,][]{GN05,MI08,vR08,Ed09}
illustrate the aspects of coronal reconnection that are---and are
not---modeled well by potential fields, and future studies need to
account for these effects more consistently.
Also, analytic models of the micro-scale kinetic physics should
be developed further in order to complement the coarser-gridded
numerical simulations.
Ideas such as stochastic growth theory \citep{CR98}
or non-modal stability \citep{Cp10} may be useful ways to understand
the partitioning of energy within reconnection regions.

Additional work should be done to refine and test the idea that the
supergranular network is the natural by-product of smaller-scale
granular activity \citep{Ra03}.
Our success in reproducing the measured autocorrelation patterns
in magnetograms (see Figure \ref{fig05}) does not necessarily imply
that there is no convective component to supergranulation.
However, our results do appear to provide evidence that at least
{\em some} of the 10--30 Mm magnetic structure on the Sun can be
built up from $\sim$1 Mm granulation effects via a kind of
diffusion-limited aggregation \citep[see also][]{Ch07}.

Another topic that requires further study is the coupling between
waves and flux emergence in the granular convective flows at the
photospheric lower boundary.
\citet{CvB05} estimated that the surface-averaged energy flux
of Alfv\'{e}n waves in the low corona is of order $10^6$
erg cm$^{-2}$ s$^{-1}$ \citep[see Figure 12 of][]{CvB05}.
It is probably not a coincidence that this is of the same order of
magnitude as the Poynting flux $S$ due to the emergence of
ephemeral regions.
The interplay between convective overturning motions, colliding
granular cells, and thin flux tubes may give rise to a rough
equipartition between these different sources of energy.
By constructing models that contain the seeds of {\em both}
WTD and RLO processes, we can better determine their relative
contributions to coronal heating and solar wind acceleration.

\acknowledgments

The authors gratefully acknowledge Ben Chandran, Phil Isenberg,
Terry Gaetz, and the anonymous referee for valuable discussions.
This work was supported by the National Aeronautics and Space
Administration (NASA) under grant {NNX\-09\-AB27G} to the
Smithsonian Astrophysical Observatory.
The SOLIS data used in this paper are produced cooperatively
by NSF/NSO and NASA/LWS.

\appendix
\section{An Idealized Application of Longcope's MCC Model for
Anemone-Type Events}

In this section we show how the \citet{Lg96} MCC model can be
applied to the results of the BONES simulations described above.
In this model, the motions of discrete flux sources on the solar
surface give rise to stresses in the coronal field that are
concentrated at topological boundaries (i.e., separatrix
surfaces and separator field lines).
Electric currents are assumed to form along the separators, and
then dissipate as magnetic reconnection occurs in response to
the evolution of the flux domains.
\citet{Lg96} found that the power dissipated in a single flux transfer
event must be choppy and intermittent, but its time average can be
written as
\begin{equation}
  \bar{P} \, = \, \theta_{\rm L} \frac{I^{\ast}}{2c} \left|
  \frac{d\Phi}{dt} \right|  \,\, ,
  \label{eq:longpower}
\end{equation}
where $d\Phi / dt$ is the time derivative of magnetic flux that is
in the process of transferring its connectivity, $I^{\ast}$ is
a characteristic current that is assumed to flow along the separator,
$\theta_{\rm L}$ is a dimensionless threshold constant describing the
intermittency of reconnection, and $c$ is the speed of light in vacuum.

In the double-bipole configuration of \citet{Lg96}, the transfer of
magnetic flux ($d\Phi / dt$) occurred because a fraction of the flux
from the positive pole of one bipole became reconnected with the
negative pole of the other bipole.
In our model, we consider the transfer of flux from a closed flux
tube to an open flux tube, or from open to closed.
Equation (B9) of \citet{Lg96} gave the characteristic current used
in Equation (\ref{eq:longpower}) above.
Correcting a typographical error in \citet{Lg96}, this current is
given by
\begin{equation}
  I^{\ast} \, = \, \frac{c \bar{B}_{\perp}' L^{2} s}{8\pi^2}
  \label{eq:Istar}
\end{equation}
where $L$ is the length of the separator field line, $s$ is a
dimensionless geometrical constant (with $s=1$ corresponding to
a circularly shaped separator field line), and $\bar{B}_{\perp}'$
is an average value of the Jacobian-like perpendicular derivative
of the vector field at the separator,
\begin{equation}
  B_{\perp}' \, = \, \sqrt{- \mbox{det} \left( \nabla_{\perp}
  {\bf B}_{\perp} \right)}  \,\, .
  \label{eq:Jacobian}
\end{equation}
In the above, the perpendicular direction is defined relative to
the separator field line.

For a given magnetic configuration, the above equations let us
estimate the power emitted from the loss of magnetic free energy
via reconnection.
However, it would be too computationally intensive to locate and
trace all of the separator field lines during every time step of
the BONES simulation.
Thus, we aim to simplify the application of
Equation (\ref{eq:longpower}) by creating a characteristic
``building block'' for the magnetic geometry in a typical
(anemone-type) opening/closing event.
These building blocks can then be assembled together in a
statistical way to account for the total amount of evolving flux
in each time step of the Monte Carlo simulations.

\citet{Wa98} described a simple model of plume/jet events in coronal
holes that involved only three discrete flux sources: two that
form a localized bipole and a third that represents a unipolar
source of open field.
As discussed in Section \ref{sec:5.5}, the energy release that is
assumed to occur in this system happens when some of the flux in the
bipole reconnects with the unipolar region, giving rise to an equal
amount of opening and closing of flux ($f_{\rm co} = f_{\rm oc}$).
For geometric simplicity, let us assume that all three flux sources
are collinear along the $x$ axis, with a negative source in between
two positive sources.
The flux evolution occurs as the negative pole of the bipole
moves away from its original positive partner and towards the
positive source of open field.
We want to evaluate the properties of this system at a representative
time in the middle of its evolution, so let us posit an additional
symmetry; i.e., we assume that the negative pole sits at the
origin ($x=0$) and the two positive poles are both equidistant from
the origin ($x = \pm d$) and of equal positive flux.
This may be an extreme simplification, since it is known that many
details of three-dimensional null-point reconnection do depend on
whether the geometry is symmetric or asymmetric \citep{AP10}.
However, the other uncertainties in the order-of-magnitude MCC
model are probably not outweighed by this issue.

To evaluate the coronal magnetic field arising from this three-pole
system, we set the flux in the positive poles to $\Phi_{+} > 0$ and
flux in the negative pole at the origin to $\Phi_{-} < 0$.
The two free parameters that constrain the topology of the field
lines are the pole separation $d$ and the ratio of negative to
positive fluxes $m = | \Phi_{-} / \Phi_{+} |$. 
Thus, Equation (\ref{eq:Bdef}) gives
\begin{equation}
  B_{x} (x,y,z) = \frac{\Phi_{+}}{2\pi} \left\{
  \frac{x+d}{[(x+d)^{2} + y^{2} + z^{2}]^{3/2}} +
  \frac{x-d}{[(x-d)^{2} + y^{2} + z^{2}]^{3/2}} -
  \frac{mx}{[x^{2} + y^{2} + z^{2}]^{3/2}} \right\}  \,\, ,
  \label{eq:Btrix}
\end{equation}
\begin{equation}
  B_{y} (x,y,z) = \frac{\Phi_{+}}{2\pi} \left\{
  \frac{y}{[(x+d)^{2} + y^{2} + z^{2}]^{3/2}} +
  \frac{y}{[(x-d)^{2} + y^{2} + z^{2}]^{3/2}} -
  \frac{my}{[x^{2} + y^{2} + z^{2}]^{3/2}} \right\}  \,\, ,
  \label{eq:Btriy}
\end{equation}
\begin{equation}
  B_{z} (x,y,z) = \frac{\Phi_{+}}{2\pi} \left\{
  \frac{z}{[(x+d)^{2} + y^{2} + z^{2}]^{3/2}} +
  \frac{z}{[(x-d)^{2} + y^{2} + z^{2}]^{3/2}} -
  \frac{mz}{[x^{2} + y^{2} + z^{2}]^{3/2}} \right\}  \,\, .
  \label{eq:Btriz}
\end{equation}
We will consider values of the flux ratio $m$ between about 0.5 and 2.
For $m > 2$, the central source ``breaks out'' with its own open
field of negative polarity, which is a situation that we are not
considering here.

\begin{figure}
\epsscale{0.50}
\plotone{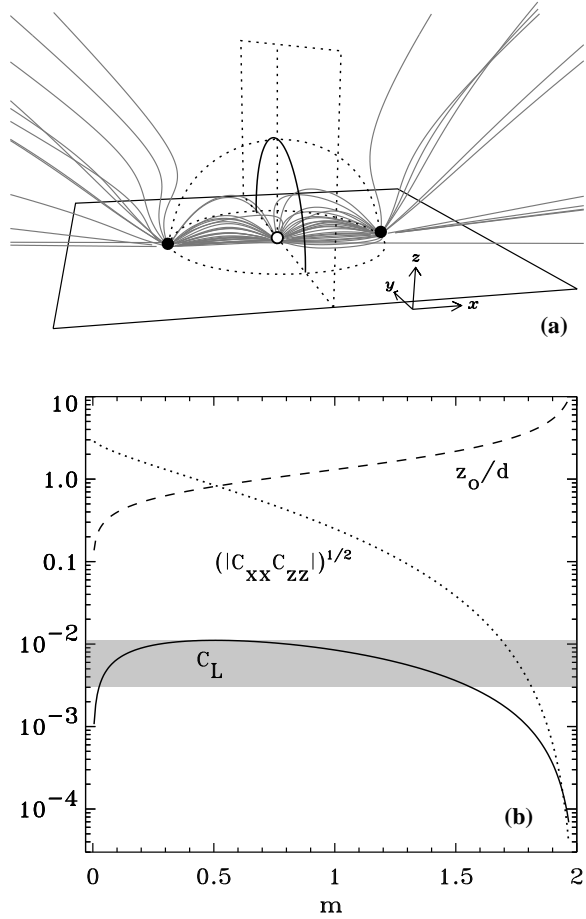}
\caption{Properties of the simple three-pole geometry used to
estimate several factors in the MCC model.
(a) Three-dimensional projection of selected field lines (gray
curves), shown along with the two positive poles (filled circles)
and the negative pole (open circle) on the surface, the separator
field line (black solid curve), and outlines of the locations of
the separatrix surfaces (dotted curves).
(b) Plot that shows how the null-point height $z_{0}/d$ (dashed curve),
the magnetic Jacobian factor $(|C_{xx} C_{zz}|)^{1/2}$ (dotted curve),
and the constant $C_{\rm L}$ (solid curve) depend on the flux
imbalance ratio $m$.
The range of values for $C_{\rm L}$ used when analyzing the results
of the BONES models is shown as a gray band.}
\label{fig13}
\end{figure}

Figure \ref{fig13}(a) illustrates a few representative field
lines for the case $m = 0.8$.
For simplicity, we assume the poles are at $z=0$.
The coronal volume ($z > 0$) can be separated into four distinct
domains according to the field-line topology:
(1) a set of open field lines that originates from the left-hand
positive pole,
(2) a set of open field lines that originates from the right-hand
positive pole,
(3) a set of closed field lines that connects the left and center
poles, and
(4) a set of closed field lines that connects the center and right
poles.
There are two separatrix surfaces that delineate the boundaries
between these domains:
a vertical surface that spans the $y$--$z$ plane and is defined
by the condition $x=0$, and the upper half of a prolate spheroidal
surface centered on the origin.
The {\em separator field line} is the intersection of the two
separatrix surfaces, and for this model it is a semicircle in the
$y$--$z$ plane.

In order to solve Equation (\ref{eq:Jacobian}) we need to evaluate
the exact position of the separator.
First, we locate its maximum height $z_0$ by looking for the height
of the magnetic null point along the vertical line denoted by
$x=0$ and $y=0$.
We use Equations (\ref{eq:Btrix})--(\ref{eq:Btriz}) to solve for
the magnitude of the magnetic field strength, but we do not worry
about its absolute normalization.
Along the vertical line in question, $B_{x} = B_{y} = 0$, and
we find that
\begin{equation}
  B_{z} \, \propto \, \frac{2}{(d^{2} + z^{2})^{3/2}} -
  \frac{m}{z^3}  \,\, .
\end{equation}
We set $B_{z} = 0$ and search for a nontrivial solution for $z_{0} > 0$.
This is a cubic polynomial equation, and Figure \ref{fig13}(b)
shows the numerical solution for the ratio $z_{0}/d$ as a function
of $m$.  Solutions exist only for $m < 2$.
Due to the symmetry in our assumed system, the separator field
line is confined to the plane $x=0$, and it subtends a semicircular
shape for $y \neq 0$.
Thus, the separator obeys $y^{2} + z^{2} = z_{0}^{2}$, its
length is $L = \pi z_{0}$, and we can use the geometrical factor
$s=1$ in Equation (\ref{eq:Istar}).

We estimate the average value of $\bar{B}_{\perp}'$ along the
separator by just computing its value at the maximum height
($x=y=0$, $z = z_0$).
At this point, the field's parallel direction points along the
$y$ axis, so Equation (\ref{eq:Jacobian}) can be written
\begin{equation}
  \bar{B}_{\perp}' \, = \, \sqrt{ \left|
  \frac{\partial B_x}{\partial x} \frac{\partial B_z}{\partial z} -
  \frac{\partial B_z}{\partial x} \frac{\partial B_x}{\partial z}
  \right| }  \,\, .
  \label{eq:Bperp}
\end{equation}
The cross-derivatives in the second term are found to be zero, and
it can be shown that
\begin{equation}
  \bar{B}_{\perp}' \, = \, \frac{\Phi_{+}}{\pi d^3}
  \sqrt{| C_{xx} C_{zz} |}  \,\, ,
\end{equation}
where
\begin{equation}
  C_{xx} \, = \, \frac{x-2}{(x+1)^{5/2}} - \frac{m}{2 x^{3/2}}
  \label{eq:Cxx}
\end{equation}
\begin{equation}
  C_{zz} \, = \, \frac{1-2x}{(x+1)^{5/2}} + \frac{m}{x^{3/2}}
  \label{eq:Czz}
\end{equation}
and $x = (z_{0} / d)^{2}$.
The two dimensionless factors given in Equations (\ref{eq:Cxx})
and (\ref{eq:Czz}) are related to Equation (\ref{eq:Bperp}) via
\begin{equation}
  \frac{\partial B_x}{\partial x} =
  \frac{\Phi_{+}}{\pi d^3} C_{xx}
  \,\, , \,\,\,\,\,
  \frac{\partial B_z}{\partial z} =
  \frac{\Phi_{+}}{\pi d^3} C_{zz}  \,\, .
\end{equation}
Figure \ref{fig13}(b) shows the dimensionless factor
$(|C_{xx} C_{zz}|)^{1/2}$ as a function of the flux imbalance ratio $m$.

The above model gives us the ability to write the average power
dissipated (Equation (\ref{eq:longpower})) as
\begin{equation}
  \bar{P} \, = \, \theta_{\rm L} C_{\rm L} \frac{\Phi_{+}}{d}
  \left| \frac{d\Phi}{dt} \right|  \,\, ,
  \label{eq:avgpower}
\end{equation}
where the dimensionless factors dependent on $m$ have been collected
into a single constant
\begin{equation}
  C_{\rm L} \, = \, \frac{1}{16\pi} \sqrt{| C_{xx} C_{zz} |}
  \left( \frac{z_0}{d} \right)^{2}  \,\, .
\end{equation}
Figure \ref{fig13}(b) shows that $C_{\rm L}$ varies less strongly
as a function of $m$ than either of its components.
In our models, we will not keep track of the individual $m$
imbalance ratios for each reconnection event.
Instead we adopt a {\em range} of values for $C_{\rm L}$ that
spans the majority of the variation for many likely $m$ values.
The gray region in Figure \ref{fig13}(b) shows this range of values;
the lower limit is 0.003, and the upper limit is the maximum value
of 0.011.

The other dimensionless constant in Equation (\ref{eq:longpower})
is $\theta_{\rm L}$.
This parameter is a threshold ratio of the instantaneous current
density to the characteristic current $I^{\ast}$, and in the MCC
model it is assumed that plasma instabilities (e.g., the
ion-acoustic instability or tearing-mode instabilities) will
limit the growth of the current to some fraction of $I^{\ast}$.
\citet{Lg96} argued that $\theta_{\rm L} \ll 1$ was reasonable to
expect, and he ended up using $\theta_{\rm L} = 0.15$ in the
initial MCC models.
However, \citet{LS98} and \citet{LK99} found that some situations
appear to demand larger values of order $\theta_{\rm L} \approx 1$.
We will use the latter value, but we will keep in mind that the
resulting heating rate may be an upper limit.

To apply the heating rate derived above to our Monte Carlo models,
we note that Equation (\ref{eq:dBdtco}) gives the time derivative
of magnetic flux that is being opened up in the simulation box,
during each time step.
In order to solve Equation (\ref{eq:avgpower}), however, we also
need to know the characteristic fluxes in the elements that are
interacting, as well as their inter-element distances.
Since many reconnection events may be occurring simultaneously
in each time step, we must use {\em averages} taken over the box
area.
We also divide both sides of Equation (\ref{eq:avgpower}) by $A$
in order to express the heating rate per unit area in terms of the
variations in magnetic flux density.
Thus,
\begin{equation}
  \langle F_{\rm co} \rangle \, = \,
  \theta_{\rm L} C_{\rm L} \frac{\Phi_1}{\langle d \rangle}
  \left| \frac{dB}{dt} \right|_{\rm co} 
\end{equation}
where $\Phi_{1} = \Phi_{\rm abs} / N$ is the mean absolute flux
per element in the simulation box, and
\begin{equation}
  \langle d \rangle \, = \, \sqrt{\frac{4A}{\pi N}}
\end{equation}
is the mean separation between flux elements.
This is the form of the MCC energy flux used for the BONES results
presented in Section \ref{sec:5.5}.

\end{document}